\documentclass[fleqn,usenatbib]{mnras}

\usepackage{newtxtext,newtxmath}

\usepackage[T1]{fontenc}
\usepackage{ae,aecompl}

\usepackage{graphicx}	
\usepackage{amsmath}	
\usepackage{natbib}     
\usepackage{longtable}  
\usepackage{CJK}        

\newcommand{\n}{NGC~}

\newcommand{\Ha}{H$\alpha$}

\newcommand{\HI}{\text{H\,\textsc{I}}}
\newcommand{\MgII}{\text{Mg\,\textsc{II}}}
\newcommand{\msun}{$M_{\odot}$}
\newcommand{\um}{~$\mu$m}

\newcommand{\logMdust}{\rm log ($M_{\rm dust}/M_{\odot}$)}

\newcommand{\jhy}{}

\newcommand{\be}{\begin{equation}} \newcommand{\ba}{\begin{eqnarray}}
\newcommand{\ee}{\end{equation}} \newcommand{\ea}{\end{eqnarray}}

\def\-{{\em{---}}}

\def \h         {\hbox{$\, h$} }

\def\H7{\mbox {$h_{0.7}$}}

\def\IZw18{I~Zw~18}
\def\m82{M82}

\def\h{\mbox {\rm H}}

\def\msun{\mbox {${\rm ~M_\odot}$}}

\def\lsun{\mbox {${~\rm L_\odot}$}}
\def\msunyr{\mbox {$~{\rm M_\odot}$~yr$^{-1}$}}

\def\Ha{\mbox {H$\alpha$}}

\def\h0{\mbox {~H$_0$}}
\def\q0{\mbox {~q$_0$}}
\def\o3hb{[OIII]$\lambda5007$~/~H$\beta$~}
\def\O1ha{[OI]$\lambda6300$~/~H$\alpha$~}

\def\s2ha{[SII]$\lambda\lambda6717,31$~/~H$\alpha$~}
\def\2z2{HeII~$\lambda4686$~}
\def\z7{[NII]~$\lambda6583$ }
\def\N2{[NII]~$\lambda6583$~/~H$\alpha$~}
\def\16z2{[SII]~$\lambda\lambda6717, 6731$ }

\def\n{NGC~}
\def\asec{\ifmmode {'' }\else $''~$\fi}  
\def\amin{\ifmmode {' }\else $'~$\fi}    
\def\arcsper{\ifmmode \rlap.{'' }\else $\rlap{.}'' $\fi} 
\def\arcmper{\ifmmode \rlap.{' }\else $\rlap{.}' $\fi} 
\def\sles{\lower2pt\hbox{$\buildrel {\scriptstyle <}
   \over {\scriptstyle\sim}$}} 
\def\sgreat{\lower2pt\hbox{$\buildrel {\scriptstyle >}
    \over {\scriptstyle\sim}$}} 
\def\kms{\mbox {~km~s$^{-1}$}}

\def\flux{~ergs~s$^{-1}$~cm$^{-2}$}

\def\cm3{~cm$^{-3}$}

\def\mpc3{~Mpc$^{3}$}
\def\mpc-3{~Mpc$^{-3}$}

\def\um{~${\mu}$m}

\def\apj{ApJ}
\def\apjs{ApJS}
\def\pasp{PASP}
\def\aj{AJ}

\def\nat{Nature}

\def\pasj{PASJ}
\title[Dust Content of Galactic Halos. III. \n891.]{Exploring the Dust Content of
  Galactic Halos with {\it Herschel} III. \n891}

\author[J. H. Yoon et al.]{
J. H. Yoon,$^{1}$ 
C. L. Martin,$^{1}$ \thanks{E-mail: cmartin@.ucsb.edu}
S. Veilleux,$^{2,3}$
M. Mel\'{e}ndez,$^{2,4}$
T. Mueller,$^{5}$
\newauthor
K. D. Gordon,$^{4}$
G. Cecil,$^{6}$
J. Bland-Hawthorn,$^{7}$
C. Engelbracht$^{8}$
\\
$^{1}$Department of Physics, University of California, Santa Barbara, CA 93106, USA\\
$^{2}$Department of Astronomy, University of Maryland, College Park, MD 20742, USA\\
$^{3}$Joint Space-Science Institute, University of Maryland, College Park, MD 20742, USA\\
$^{4}$Space Telescope Science Institute, Baltimore, MD 21218, USA\\
$^{5}$Max-Planck-Institute for Extraterrestrial Physics (MPE), 85748 Garching, Germany\\
$^{6}$Department of Physics and Astronomy, University of North Carolina, Chapel Hill, NC 27599\\
$^{7}$Department of Physics, University of Sydney, Sydney, NSW 2006, Australia\\
$^{8}$Department of Astronomy, University of Arizona, Tucson, AZ 85721, USA
}

\date{Accepted XXX. Received YYY; in original form ZZZ}

\pubyear{2020}

\begin{document}
\label{firstpage}
\pagerange{\pageref{firstpage}--\pageref{lastpage}}
\maketitle

\begin{abstract}
We present deep far-infrared observations of the nearby edge-on galaxy
\n891 obtained with the {\it Herschel Space Observatory} and the {\it
Spitzer Space Telescope}. The maps confirm the detection of thermal
emission from the inner circumgalactic medium (halo) and spatially resolve 
a dusty superbubble and a dust spur (filament).
The dust temperature of the halo component is lower than that of the
disk but increases across a region of diameter $\approx 8.0$
kpc extending at least 7.7 kpc vertically from one side of the
disk, a region we call a superbubble because of its association
with thermal X-ray emission and a minimum in the synchrotron scaleheight.
This outflow is breaking through
the thick disk and developing into a galactic wind, which is of
particular interest because \n891 is not considered a starburst galaxy; 
the star formation rate surface density, 0.03\msunyr\ kpc$^{-2}$, and gas
fraction, just $10\%$ in the inner disk, indicate the threshold for
wind formation is lower than previous work has suggested. We conclude
that the star formation surface density is sufficient for superbubble
blowout into the halo, but the cosmic ray electrons may play a critical 
role in determining whether this outflow develops into a fountain or escapes
from the gravitational potential. The high dust-to-gas ratio in the dust spur 
suggests the material was 
pulled out of \n891 through the collision
of a minihalo with the disk of NGC 891.
We conclude that \n891 offers an example of both feedback and satellite interactions
transporting dust into the halo of a typical galaxy.

\end{abstract}




\section{Introduction}

Studying giant spiral galaxies beyond the Local Group determines whether the assembly of the
Milky Way was typical. Having a vantage point outside the system also makes observations of
the stellar halo and circumgalactic medium (CGM) easier. The disk component can be completely separated
from the halo,  for example,  when such galaxies are viewed edge-on.  The edge-on orientation of
the nearby galaxy, \n891, is ideal for studying the interaction between the disk, CGM, and satellite 
galaxies.

\n891 resides in a spiral-rich group representative of 
typical environments and has a luminosity similar to the group's brightest member, \n1023. 
\n891 is similar to the Milky Way in spiral type and rotation speed \citep{Swaters1997}. In both galaxies, 
a large fraction of the halo stars have been accreted from satellites \citep{Mouhcine2010,Bell2008} and recent star formation is elevated (compared to the disk as a whole) in a molecular ring and 
central circumnuclear disk \citep{Scoville1993a}.

The warm ionized gas in the interstellar medium of \n891 is more extended than the Reynold's layer \citep{Reynolds1989}
in the Milky Way, and the central surface density of the warm ionized medium is roughly twice as high
in \n891 as in the Milky Way \citep{Rand1990,Dettmar1990}. The molecular 
gas mass of \n891 is 2.5 times larger than that of the Milky Way, 
and the current star formation rate (SFR) is proportionally higher. Star formation in 
\n891 is  more active than in the Milky Way disk yet not high enough to be considered a starburst.
The specific SFR and stellar mass of \n891 are similar to many COS Halos galaxies
at redshift $z \approx 0.2$ \citep{Tumlinson2011,Werk2014a}. 
The proximity of \n891, however,
has made it possible to directly detect emission from the circumgalactic medium (CGM).  The recent
21-cm detection of cold gas out to 90 - 120 kpc along the minor axis \citep{Das2020},
combined with the X-ray detection of the virialized halo gas \citep{Hodges-Kluck2018} and 
the well-studied synchroton halo \citep{Schmidt2019}, together
provide excellent observational contraints on the properties of the inner CGM. 

The vertical distribution of hot gas \citep{Bregman1994}, neutral gas \citep{Swaters1997}, molecular 
gas \citep{GarciaBurillo1992}, warm dust \citep{McCormick2013a},  and stars \citep{vanderKruit1981} have been 
measured in \n891 and found to contain extraplanar components. 
A full 30\% of the HI resides in a halo \citep{Oosterloo2007a}. The thermal emission from the halo dust
indicates a larger dust mass than was originally estimated from absorption features 
\citep{Howk1997a,Howk2000a,Popescu2000,Bianchi2008,Popescu2011}. 
The first maps of the dust mass and temperature
showed a strong correlation with the distribution of stellar mass and star formation, respectively \citep{Hughes2014}. 
The X-ray halo of \n891 was the first detected in a non-starburst galaxy, but it has
remained controversial as to whether the hot gas was heated by galactic accretion \citep{Hodges-Kluck2012a}
or feedback processes \citep{Strickland2004a,Strickland2004b,Tullmann2006}. 

The presence of cold dust in \n891 has been known for several decades, but little was known about
its spatial distribution due to the low resolution of the far-infrared and submm observations. Observations of 
\n891 with {\it Spitzer}/MIPS resolved disk and halo components at 24\um\ but lacked the spatial resolution
at longer wavelengths required to map the 
spectral energy distribution, hereafter the SED
(GO-20528, PI C. Martin).
As part of the Very Nearby Galaxies Survey, a {\it Herschel} Guaranteed Time Key Project, the large vertical 
extent of the dust in \n891 was resolved with {\it Herschel} PACS/SPIRE \citep{Hughes2014}

In this paper, we combine  deeper {\it Herschel} observations (OT1\_sveilleu\_2, PI S. Veilleux)
with the {\it Spitzer}/MIPS data to better describe the amount of halo dust and its origin. 
This work builds on our  examination of \n4631 \citep[][hereafter Paper I]{Melendez2015a}, a lower mass galaxy with 
a higher SFR surface density \citep{Tullmann2006}, and six nearby dwarf galaxies including \n1569 \citep[][hereafter Paper II]{McCormick2018}. The interstellar dust in \n891 plays an important 
role in its appearance. Young, star-forming regions, for example, are much more apparent along the northeast 
(approaching) side of the disk compared to the receding (southwest) half due to the accumulation of dust
along the inner edges of the trailing spiral arms. Our deep observations directly show the relative importance
of stellar feedback and satellite interactions for building a dusty halo. This new perspective should
prove generally valuable for understanding the dust-to-gas ratio in galaxy halos \citep{Menard2012a}, and the processes taking place in the outskirts of galaxies in general \citep{Veilleux2020}.

Our presentation is structured as follows. In Section~\ref{sec:observations}, we present the far-IR/submm
observations. 
The morphological components of the dust distribution are defined in Section~\ref{sec:morph}
before describing our fits to the multi-band photometry
and the resulting physical properties in Section~\ref{sec:temp_mass}. 
In Section~\ref{sec:discuss},
we argue that the superbubble will become a galactic wind and discuss the origin of a puzzling,
extraplanar dust spur.  Our conclusions are briefly summarized in Section~\ref{sec:conclusions}.

We adopt a distance of 9.77~Mpc \citep{Ferrarese2000a}, which is the weighted mean of distances
derived from HST observations of planetary nebula and surface brightness fluctuations.
This distance gives an angular scale of 47.37~pc/\arcsec.

\section{Observation and Data Reduction} \label{sec:observations}

\subsection{Spitzer/IRAC and Spitzer/MIPS  data}
\label{spitzer.sec}

{\it Spitzer} images at 3.6\um, 4.5\um, and 24\um\ are used in this study. Images taken with 
the Infrared Array Camera (IRAC) at 3.6\um\ and 4.5\um\ cover the entire galaxy 
(PID: 3, PI: G. Fazio).  The observations consist of a series of
dithered images taken with exposure times from 0.4 to 96.80
seconds.  
The basic calibrated
data were obtained from the {\it Spitzer} archive and combined into a 
mosaic using the
{\sc montage} 
package (http://montage.ipac.caltech.edu/) correcting for
variations in the overall level and rejecting outliers.  The point
source flux accuracy is $\sim$2\% \citep{Reach2005} and the extended
source surface brightness accuracy is estimated at $\sim$4\%.

We observed \n891 with the {\it Multiband-Imaging Photometer
for Spitzer} \citep[MIPS,][]{Rieke2004a} at 24\um\ 
(PID: 20528, PI: C. Martin). 
The images were obtained using
the medium rate, scan-map mode and the scan legs of $0.5^\circ$ with
148\arcsec\ cross-scan offsets. This observation consists of 5 scans
(Obs. id 14815488, 14815744, 14816000, 14816256, 14816512) with total
exposure times of 1600 seconds. The data were reduced with the MIPS
Data Analysis Tool v3.04 \citep[DAT][]{Gordon2005a}.  Extra steps were
carried out to improve the images including readout offset correction,
array averaged background subtraction (using a low order polynomial
fit to each leg, with the region including NGC 891 excluded from this
fit), and exclusion of the first five images in each scan leg due to
boost frame transients.  The point source flux accuracy is $\sim$2\%
\citep{Engelbracht2007} and the extended source surface brightness
accuracy is estimated at $\sim$4\%.

The 24\um\ image resolves  stars and background galaxies, which
we mask out for accurate photometry of the galaxy. 
We identified background galaxies using \texttt{SExtractor}. Briefly, we
set the detection threshold to 1.5$\sigma$ in surface brightness. The sources
identified by the segmentation map were then replaced by the values in the
background map at the same locations.

%

\subsection{Herschel/PACS and Herschel/SPIRE Data}

The Photodetector Array Camera and Spectrometer \citep[PACS,][]{Poglitsch2010a}
observation for NGC 891 was conducted as our cycle 1 open-time program
(OT1\_sveilleu\_2, PI: S. Veilleux). This program utilizes PACS blue
70\um\ and red 160\um\ channels with 6 scan position angles at 55$^{\circ}$,
70$^{\circ}$, 85$^{\circ}$, 95$^{\circ}$, 110$^{\circ}$, and 125$^{\circ}$
(Obs IDs: 1342237999, 1342238000, 1342238001, 1342238002, 1342238003, and
1342238004). The exposure time per each scan was 5596 seconds (total 9.3
hours) including overhead. 
 The scan map leg length and separation were
4\arcmin\ and 4\arcsec\ and the number of scan map legs was 60. Each scan map was repeated three times for redundancy reasons.

We also retrieved, reduced, and analyzed the PACS data from {\it The Herschel EDGE-on galaxy Survey}
\citep[HEDGES,][OT2\_emurph01\_3]{Murphy2011a} for the analysis of extended
structure as these data have a larger field-of-view than do our observations. 
For the blue channel
(70\um), the scan angle was 45$^{\circ}$ and 135$^{\circ}$ with an exposure
time of 6550 seconds. The scan map leg length and separation were 14\farcm6
and 14\farcs0 and the number of scan map legs was 70. For the green channel
(100\um), the exposure time was 8977 seconds at scan angles of 45$^{\circ}$
and 135$^{\circ}$. The scan map leg length and separation were 14\farcm6
and 20\farcs0 and the number of scan map legs was 51.

The observation in 250\um, 350\um, and 500\um\ bands were performed using
 the Spectral and Photometric Imaging Receiver \citep[SPIRE,][]{Griffin2010a}
onboard {\it Herschel Space Observatory}. Our SPIRE images are the part of the
KINGFISH open-time key program, KPGT\_cwilso01\_1, from {\it Herschel} guaranteed
time key project,{\it Physical Processes in the Interstellar Medium of Very
Nearby Galaxies}\footnote{http://hedam.lam.fr/VNGS/index.php} \citep[ID:
1342189430,][]{Griffin2010a}. The beam sizes are 465.39, 822.58, and 1768.66
arcsec$^2$ at 250, 350, and 500\um, respectively (SPIRE Data Reduction
Guide, Table 6.7) and the exposure times were 1678 seconds for each channel.

\subsection{Data Reduction}

For the PACS data, we used the {\it Herschel Interactive Processing Environment}
\citep[HIPE,][]{Ott2010a} version 10.0 for data reduction. This reduction
procedure contains the extraction of the calibration tree for the data
processing, electronic crosstalk correction, flat-fielding, and unit conversion
from volts to Janskys per pixel. The final map was created with the algorithm
implemented in {\it Scanamorphos} \citep[v24.0,][]{Roussel2013a}. The error
and weight maps were also produced with {\it Scanamorphos} and the error
map is defined as the error on the mean brightness in each pixel. 

The SPIRE data were processed from level 0 up to level 1 with HIPE scripts
in the Scanamorphos distribution. The same steps as PACS processing were
conducted. The thermal drifts are subtracted by using the smoothed series
of thermistors located on the detector array as the input of the drift model (see \citet{Ott2010a} for details).
A more detailed description of the data reduction can be found in Paper I.

Both sets of PACS observations contain a
low-amplitude artifact parallel to the major axis on the eastern side of the disk midplane. It is 70\arcsec\ away to the south-east and the flux in the ghosts is less than 0.5\% of the corresponding object flux.
This region immediately east of the disk midplane was not used in our analysis of the vertical structure of the disk.
The amplitude of the feature is small enough, that it has a negligible effect on our integrated flux measurements.



\begin{figure*}
\begin{center}
 \includegraphics[width=\linewidth]{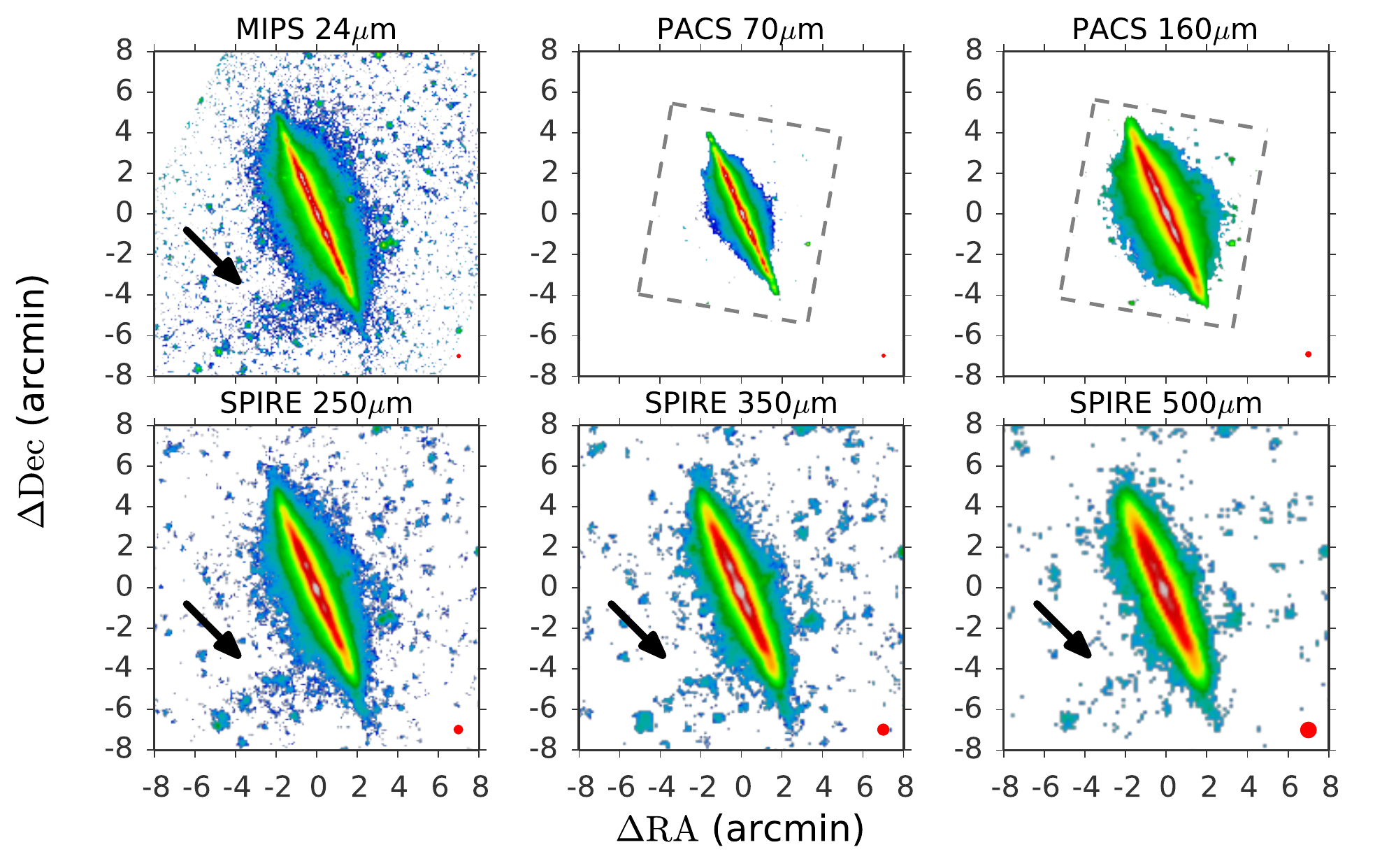}
\caption{{\it Spitzer} MIPS 24\um, {\it Herschel} PACS 70 and 160\um, and
  {\it Herschel} SPIRE 250, 350, and 500\um\ images of \n891 The color bars
  show surface brightness on a logarithmic scale; the maximum central
  surface brightness is white, and the images are grayed out below a
  S/N ratio of 1.5. The black arrow indicates the dust spur to the
  south-east. The red circles indicate the size of the point spread
  function. 
The color bar ranges (in Janskys per pixel on a logarithmic scale)
are as follows:  0.0032 to 63 (24 micron), $10^{-5}$ to 0.079 (70 micron), 
$10^{-5}$ to 0.40 (160 micron), $10^{-4}$ to 0.50 (250 micron),
$10^{-4}$ to 0.25 (350 micron), and 
$10^{-4}$ to 0.16 (500 micron).
}
\label{fig:6bands}
\end{center}
\end{figure*}

\subsection{Integrated Photometry of \n891 \& Global Properties}

Figure~\ref{fig:6bands} compares the new (24, 70, 160\um) images to
the SPIRE images at 250, 350, and 500\um. We define the center
of \n891 by the location of the maximum 4.5\um\ IRAC intensity, a proxy
for the center of the stellar bulge. In Figure~1 and subsequent images, 
the coordinates (0,0) reference this position. Table~\ref{tab:ngc891} provides the 
corresponding sky coordinates. 

We measured fluxes from the  24, 70, 160, 250, 350, and 500\um~images.
In order to match the point spread function (PSF), we convolved the images with the kernels provided
 by \citet{Aniano2011a}. The image pixel scale and coordinates were then
matched using the IDL code \texttt{hrebin.pro} and \texttt{hastrom.pro}
in the Goddard package. These processes conserve total image flux. 

\begin{figure}
 \begin{center}
  \includegraphics[width=\linewidth, angle=0, trim = 0 0 0 0]{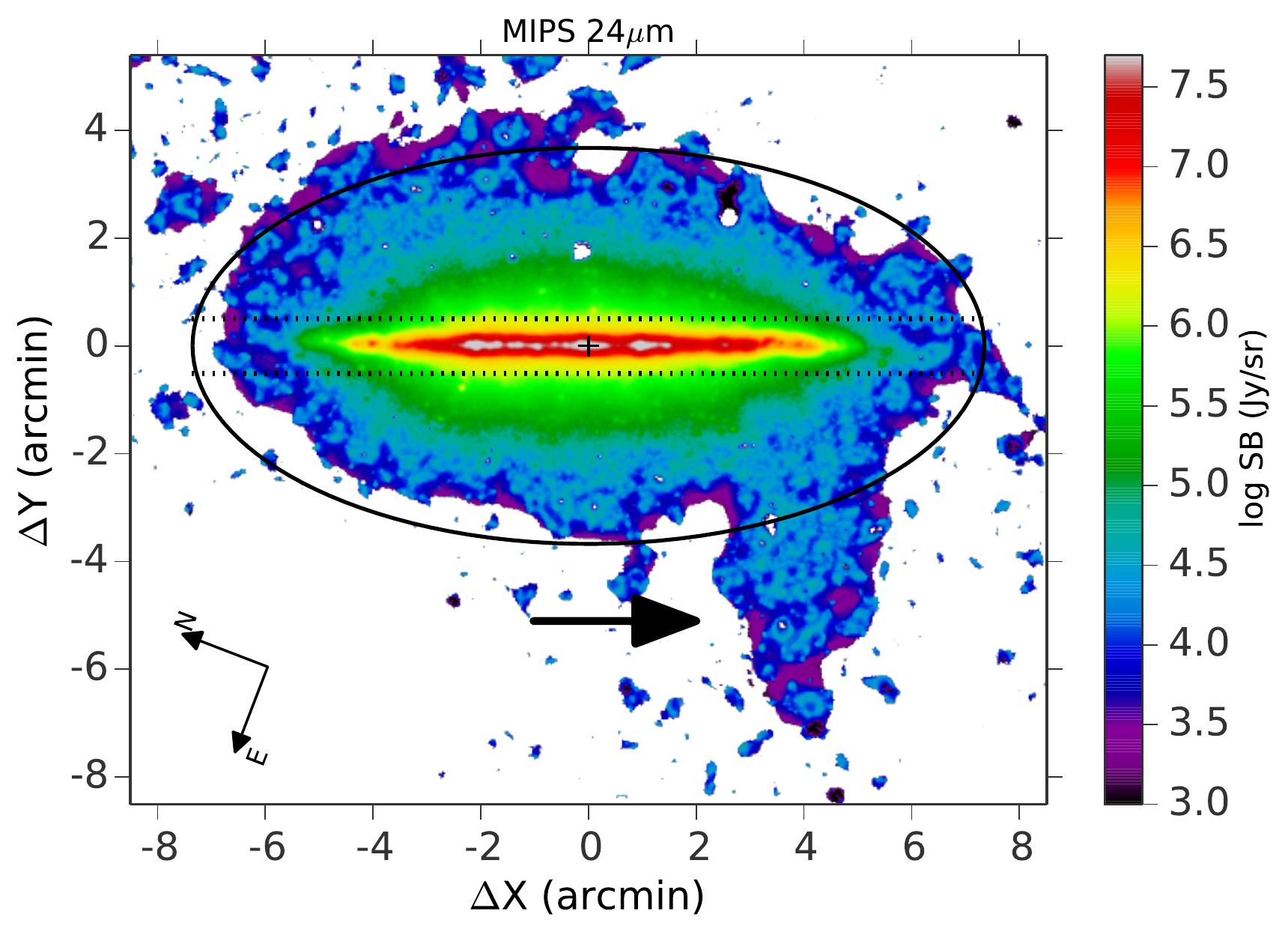}
   \caption{{\it Spitzer}/MIPS 24 \um\ image for NGC 891 adaptively smoothed after subtraction of background sources.
     The color scale for the surface brightness is logarithmic. 
     The scale height and scale length of the stars are indicated by the rectangle, 
     $\pm0\farcm5$ (1.4 kpc)  by  $\pm 7\farcm35$ (20.9 kpc) \citep{Ibata2009a}.
     The extended emission forms a halo which we define by an ellipse 
     (semi-major axis of $7\farcm35$ and semi-minor axis of 3\farcm68, or 10.5~kpc).
     An arrow indicates the dust spur.}
  \label{fig:clean24}
  \end{center}
\end{figure}

The aperture adopted for the integrated photometry is based on a growth curve measured 
from photometry of the SPIRE 500\um~image in increasingly larger apertures. It is an 
elliptical aperture of semi-major axis of 7\farcm35 and semi-minor axis of
3\farcm68 that includes most of the dust emission seen in 24\um\ image after
convolution to the PSF of the 500\um\ image. Figure~\ref{fig:clean24} shows this
aperture on the 24\um~image.

Within an  annulus at larger radii, we masked point sources and background galaxies 
based on the morphology in the 24\um\ data. The sky background was taken as
the mean data value following 2$\sigma$ clipping. The variance in this region
was used to compute the standard deviation of the mean. This uncertainty on
the mean background dominates the total background error for large apertures.
We also include a term for the pixel-to-pixel statistical variance in the background.

The error budget is dominated by the calibration uncertainties, which
exceed the uncertainties from both background subtraction and photon noise.
Following Paper I, we conservatively estimate the background error at 10\% in all bands.





We computed color corrections for the PACS and SPIRE fluxes assuming a
blackbody spectrum modified by the dust emissivity, $\kappa_{\nu}=\kappa_{\rm 0} (\nu/\nu_{\rm 0})^{\beta}$
with $\beta = 2$, and a blackbody temperature of $T = 30$~K.\footnote{
 As the PACS and SPIRE pipelines compute fluxes assuming a monochromatic spectrum, we applied a color correction.
  See section 4.3 in ``PACS Photometer Passbands and Color Correction
  Factors for Various Source SEDs'', PICC-ME-TN-038, version 1.0, and
  Table 5.7 in SPIRE Handbook, HERSCHEL-DOC-0798, version 2.5.}
We divided the original flux measurements  by the following correction factors:
0.977, 1.037, 0.9796, 0.9697, and 0.9796 for, 
respectively, 70\um, 160\um, 250\um, 350\um, and 500\um. 
The estimated aperture corrections were small compared to the photometric errors and were therefore not applied.

The integrated fluxes listed in Table~\ref{tab:ngc891} for the 24, 70, 160, 250, 350,
and 500\um~bands.
Our measurements
agree  with  both \citet{Bendo2012a} and \citet{Hughes2014}
to within the $1\sigma$ errors. 


\begin{table}
	\centering
	\caption{Properties of \n891}
	\label{tab:ngc891}
	\begin{tabular}{lr} %
		\hline
		Property & Value \\
		\hline
RA  (J2000)$^{a}$             & 2:22:33.0 \\
DEC (J2000)$^{a}$             & 42:20:53 \\
d (Mpc)$^{b}$                 & 9.77 \\
$V_{\rm circ}$ (\kms)$^{c}$       & 225 \\
SFR (\msunyr)                   & 5.0 \\
$F_{\nu}^{ 24\mu m}({\rm Jy})$  & $ 6.2 \pm 0.6$ \\
$F_{\nu}^{70\mu m}({\rm Jy})$   & $ 100  \pm 10$ \\
$F_{\nu}^{160\mu m}({\rm Jy})$  & $ 334 \pm 33$ \\
$F_{\nu}^{250\mu m}({\rm Jy})$  & $ 167 \pm 17$ \\
$F_{\nu}^{350\mu m}({\rm Jy})$  & $ 72 \pm 7$ \\
$F_{\nu}^{500\mu m}({\rm Jy})$  & $ 26 \pm 3$ \\
		\hline
	\end{tabular}
\\ $^{a}${Coordinates of the maximum intensity in the 4.5\um\ IRAC
  image.} $^{b}${Distance from \citet{Ferrarese2000a}.}
$^{c}${Rotation speed from \citet{Swaters1997}.}
\end{table}

\begin{figure*}
\begin{center}
  \includegraphics[trim={0 0 0 0}]{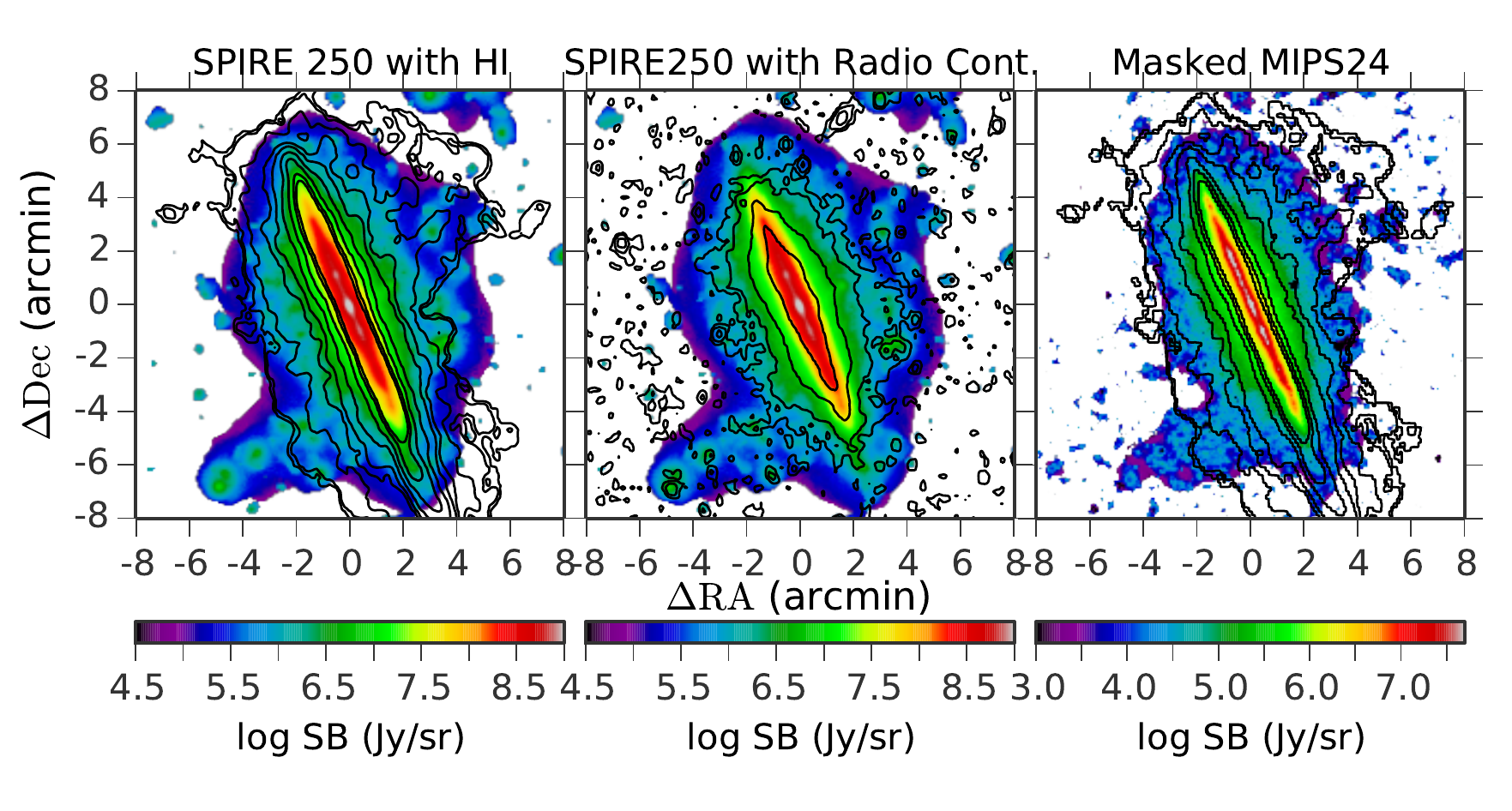}
\caption{Adaptively smoothed maps.
We applied  \texttt{ADAPTSMOOTH} \citep{Zibetti2009a} with a minimum S/N = 5  pixel$^{-1}$.
{\it Left:} The {\it Herschel} 250\um\ map overlaid with the \HI~emission contours from \citet{Oosterloo2007a}.
The contour levels for \HI~emission  are 0.01, 0.03, 0.08, 0.16, 0.47, 
0.78, 1.56, 3.11, 6.23, 9.34$\times \rm 10^{21}~cm^{-2}$. 
{\it Middle:} The {\it Herschel} 250\um\ map overlaid with the radio continuum contours
from \citet{Oosterloo2007a}. The contour levels for the radio continuum are
0.1, 0.4, 2, and 10 mJy beam$^{-1}$.
%
{\it Right:}  The {\it Spitzer} 24\um\ map with the \HI\ contours. The discrete sources in the background
were masked prior to the smoothing.
}
\label{fig:spire_smooth}
\end{center}
\end{figure*}

\begin{figure*}
\begin{center}
  \includegraphics[width=\linewidth]{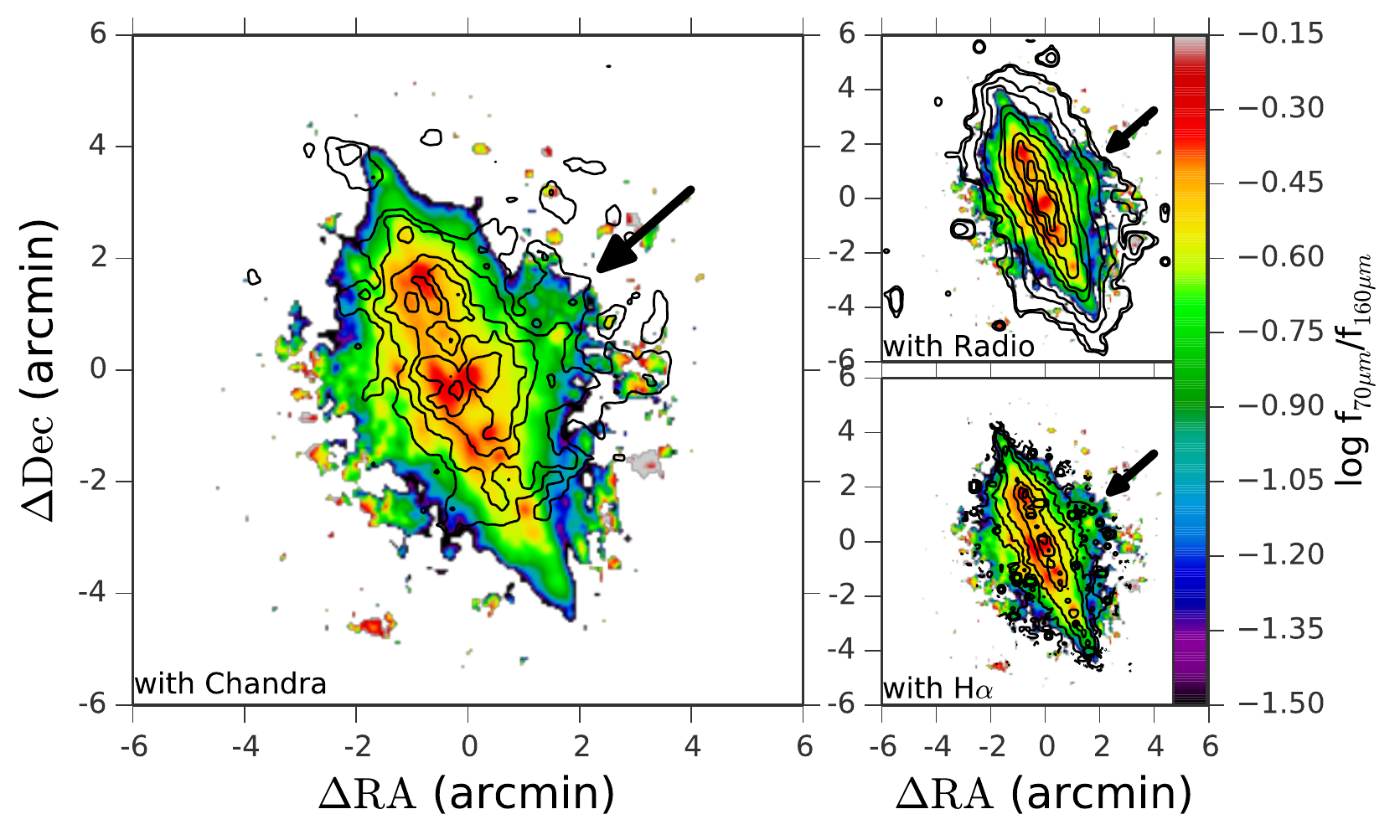}
\caption{The PACS 70\um/160\um\ color map.
The arrow points at the western superbubble.
{\it Left:}
The X-ray contours for the adaptively smoothed {\it Chandra} image 
(0.3-2.0 keV) presented by \citet{Hodges-Kluck2012a}; the contour 
levels are 9, 10, 15, 20, 30, 40, and 50 counts s$^{-1}$ deg$^{-2}$. 
{\it Top-right:}
The same color maps overlaid with radio continuum contours. The spatial resolution is 17\arcsec\ $\times$ 12\arcsec\ and the rms noise is 23 $\mu$Jy beam$^{-1}$ \citep{Oosterloo2007a}.
The contour levels for the radio continuum are
0.1, 0.4, 2, and 10 mJy beam$^{-1}$.
{\it Bottom-right:}
The same color maps overlaid with \Ha\ surface brightness contours
from an \Ha\ image taken at Steward Observatory by C. Engelbracht
and calibrated to the total luminosity reported in \citet{Hoopes1999}.
Contours are shown at $\Sigma(H\alpha) = 9.9 \times 10^{-15},
1.0 \times 10^{-14},
1.23\times 10^{-14},
2.61 \times 10^{-15},
6.09 \times 10^{-15},
1.02 \times 10^{-13}$\flux\ arcsec$^{-2}$.
}
\label{fig:pacs_color}
\end{center}
\end{figure*}

\section{Morphological Features of the Warm Dust in \n891} \label{sec:morph}


The 24\um~image in Figure~\ref{fig:clean24} resolves several structures
contributing to the thermal emission.  The brightest of these are thin
and thick dust components aligned with the center of the galaxy, components
we will refer to as the dusty disk and  dusty halo, respectively. Modeling
these components in the longer wavelength {\it Herschel} images is 
sensitive to how the PSF is modeled \citep{Bocchio2016}.

We adopt the scale heights that \citet{Bocchio2016} fitted to the
vertical surface brightness profiles. Figure 2 of their paper
shows that the shape of the vertical surface brightness profile 
requires a two-component fit. The thin, unresolved dust component
significantly reduces the scale height of the fitted thick dust component.
A consequence of this interplay is that the thick dust component 
has the largest 24\um\ scale height above the region of the stellar disk where 
the thin dust component  is faintest. Based on the vertical profile
perpendicular to a region of the disk with low star formation activity,
\citet{Bocchio2016} conclude that the scale height of the dusty halo
is $h_z = 1.36 \pm 0.01$~kpc.
The scale height of the dusty halo is therefore comparable to that of the thick
stellar disk, whose size  we have indicated in Figure~\ref{fig:clean24}. 



Our analysis describes two non-axisymmetric components of the far-infrared morphology:
(1) the large filament marked by the arrow in Figure~\ref{fig:clean24}, and (2)  a
dusty superbubble erupting from the center of the galaxy. The superbubble was
detected previously at X-ray energies \citep{Hodges-Kluck2012a,Hodges-Kluck2018}, and
we present the first description of its dust content. The filament, or {\it dust spur},
is a new discovery enabled by the spatial resolution and sensitivty of the new observations.



\subsection{The Dust Spur}
The arrow  in Fig.~\ref{fig:clean24} indicates a filament of 24\um\ emission extending 
6\farcs90 (19.6 kpc) southeast of the dust halo. We discovered this filament 
in the {\it Spitzer} MIPS data and confirmed it using {\it Herschel}.
Figure~\ref{fig:6bands} shows the SPIRE detections. The structure is
outside the field-of-view of our new PACS images, so we examined the 
PACS images obtained previously by {\it The Herschel EDGE-on galaxy Survey} 
\citep[HEDGES,][]{Murphy2011a} and found a clear detection at at 160\um,  
a few clumps of emission at  100\um, and weaker emission at 70\um. The
far-infrared filament is therefore not an instrumental artifact.

We looked for a background cluster of galaxies at this location but found
none in either X-ray images or optical images.\footnote{The $z=0.0184$ cluster Abell~347 lies 
  to the southeast at 02:25:50.9,+41:52:30 (NED).}  Our photometry indicates a
broad spectral energy distribution, which we show is consistent with thermal 
dust emission in Section~\ref{sec:temp_mass}.  The far-infrared filament may
therefore be a component of the dusty halo of \n891, and we will refer to 
this morphological feature as {\it the dust spur}.  Spectroscopy of an emission 
line would confirm its association with \n891; alternatively, higher-resolution 
mapping would be useful to definitively rule out an unresolved galaxy cluster.

The morphology of the dust spur shows no connection to the center of the galaxy,
making an association with a galactic outflow unlikely. When projected on the sky,
the long-axis of the dust spur is roughly perpendicular to the disk. The filament
emerges at a galactocentric radius $R$ =  4\farcm0 (11 kpc) and extends 6\farcm9 (19.6 kpc) 
to the southeast, and has a width of $\sim$ 2\farcm0 (6 kpc).

The dust spur is {\it not} coincident with the large filament of neutral hydrogen
described by  \citet{Oosterloo2007a} \citep[see also][]{Pingel2018}. In Figure~\ref{fig:spire_smooth}, we show their
\HI\ 21-cm emission contours on the {\it Herschel} and {\it Spitzer} images. The
lowest contour at $N(HI)= 1\times10^{19}\rm~cm^{-2}$ extends 22~kpc
northwest of \n891.  The dust spur extends to the southeast, in contrast, and is not
visible in the \HI\ contours. 
We will show that the \HI\ filament and the dust spur have different compositions 
in Section~\ref{sec:temp_mass}. 

The middle panel of Fig.~\ref{fig:spire_smooth} compares the dust emission to the
radio continuum. The overall shape of the radio continuum contours is similar
to the distribution of far-infrared emission, whereas the \HI\ contours show
a lower ratio of vertical height to galactocentric radius. 


A close inspection of the dust spur suggests a spatial coincidence between
the brightest emission regions at 250\um\ and the local maxima in the
the faintest radio continuum contours.  This correlation should be interpreted
cautiously.  Unresolved sources in the SPIRE map could be background sources
unrelated to \n891, and the radio detections have low signal-to-noise. If
the radio sources are at the redshift of the dust spur, then
the radio flux levels of the clumps, 0.2 mJy beam$^{-1}$,
combined with the fluxes listed in Table~\ref{tab:fir_radio},
are consistent with far-IR -- radio relation defined by star-forming regions 
\citep{Yun2001a,Bell2003a}. These data raise the possibility that
the dust spur is a region of active star formation.

\subsection{The Superbubble}

We smoothed the 70 \um\ image to match the resolution of the 160 \um\ image, and then we divided the
smoothed 70 \um\ image by the  160 \um\ image to produce a color map from the PACS data. This color 
map resolves a broad plume of thermal dust emission, indicated by the arrow in Figure~\ref{fig:pacs_color}. 
This structure emerges from a region of the disk within roughly 1\farcm8 (5.1 kpc) of the nucleus, slightly
inside the  3\farcm1 radius of the molecular ring.  It extends northwest, roughly perpendicular to the disk,
reaching a height of 2\farcm7 (7.7 kpc). We refer to this structure as {\it the superbubble} but will argue 
in Section~\ref{sec:discuss} that a galactic wind is actually developing here. We will show that this region 
has a slightly higher dust temperature than the surrounding halo in Section~\ref{sec:temp_mass}. 


Contours of the soft X-ray emission \citep{Hodges-Kluck2012a} outline the perimeter of the dusty superbubble.
In  deeper X-ray observations, this hot ($kT = 0.71 \pm 0.01$ keV) gas is concentrated near the most 
active region of star formation and coexists with gas near the virial temperature, $kT = 0.20 \pm 0.01$ keV
\citep{Hodges-Kluck2018}. Harder diffuse X-ray emission is detected within 5 kpc of the galactic center; 
its physical origin remains unclear. 

The soft X-ray emission from galactic winds is produced mostly at the interface between the hot wind and a cooler
component of the multi-phase outflow \citep{Strickland2004a,Strickland2004b}. 
The smaller panels in Figure~\ref{fig:pacs_color} overlay the radio continuum contours and the \Ha\ contours
on the PACS color map. The \Ha\ image shows one prominent filament extending well into the dusty bubble, but the \Ha\
image is not sensitive enough to determine the amount of correlation with the superbubble morphology at other wavelengths.
We have shown in Figure~\ref{fig:spire_smooth} that the radio contours closely follow the morphology of
the dusty halo. As would therefore be expected, the radio contours do not describe the PACS color map well. 

The fate of the superbubble will be determined in large part by the CGM.
The inner CGM of \n891 contains much more cold gas than hot, virialized gas \citep{Hodges-Kluck2018,Das2020}

\subsection{The Dusty Halo} \label{sec:halo}

On the eastern side of the disk, the 24\um\ emission 
extends 2\farcm03 (5.76 kpc) over most of the disk, 
reaching a larger distance along the dust spur  (Fig.\ \ref{fig:spire_smooth}).
On the western side of the \n891 disk, the dusty halo is detected out
to 3\farcm68 (10.5 kpc) in the 24\um\ map.

The extent of the PAH emission is similar to that of the cold dust. \citet{McCormick2013a}
estimate a symmetric extent of $\pm 5\farcm0$ with a bulk height, $H_{e,PAH} = 7.1$~kpc.

\section{Dust Temperature and Mass} \label{sec:temp_mass}

We can derive the dust mass, $M_{\rm d}$, from the opacity per unit dust mass 
and the optical depth at far-infrared wavelengths. We fit a simple, 
single-temperature modified blackbody (MBB) function to the photometry in the 
wavelength range $100 \le\ \lambda (\mu {\rm m}) \le\ 500$. The dust 
spectrum consists of a blackbody spectrum at the dust 
temperature, $B_{\nu}(T_d)$, modified by the dust opacity, $\kappa_{\nu}$, 
such that \begin{eqnarray}
F_{\nu} = {{M_{\rm d} \kappa_{\nu} B_{\nu} (T_{\rm d})} \over {D^2}}, 
\end{eqnarray}
where $D$ is the distance to the galaxy.
This simple approach yields accurate dust masses and temperatures 
provided the normalization and the spectral index of the opacity
are consistent with full dust models, which include 
the distribution of dust grain properties and a range of interstellar 
radiation fields \citep{Bianchi2013}. 
We adopt a dust opacity model, $\kappa_{\nu}=\kappa_{\rm 0} (\nu/\nu_{\rm 0})^{\beta}$. 
The normalization,  $\kappa_{\rm 0} = 1.92~\rm cm^2~g^{-1}$ at 350\um, 
and spectral index, $\beta = 2.0$, are based on the Milky Way model presented in 
\citet{Draine2007a}.



\begin{figure}
\begin{center}
 \includegraphics[width=\columnwidth]{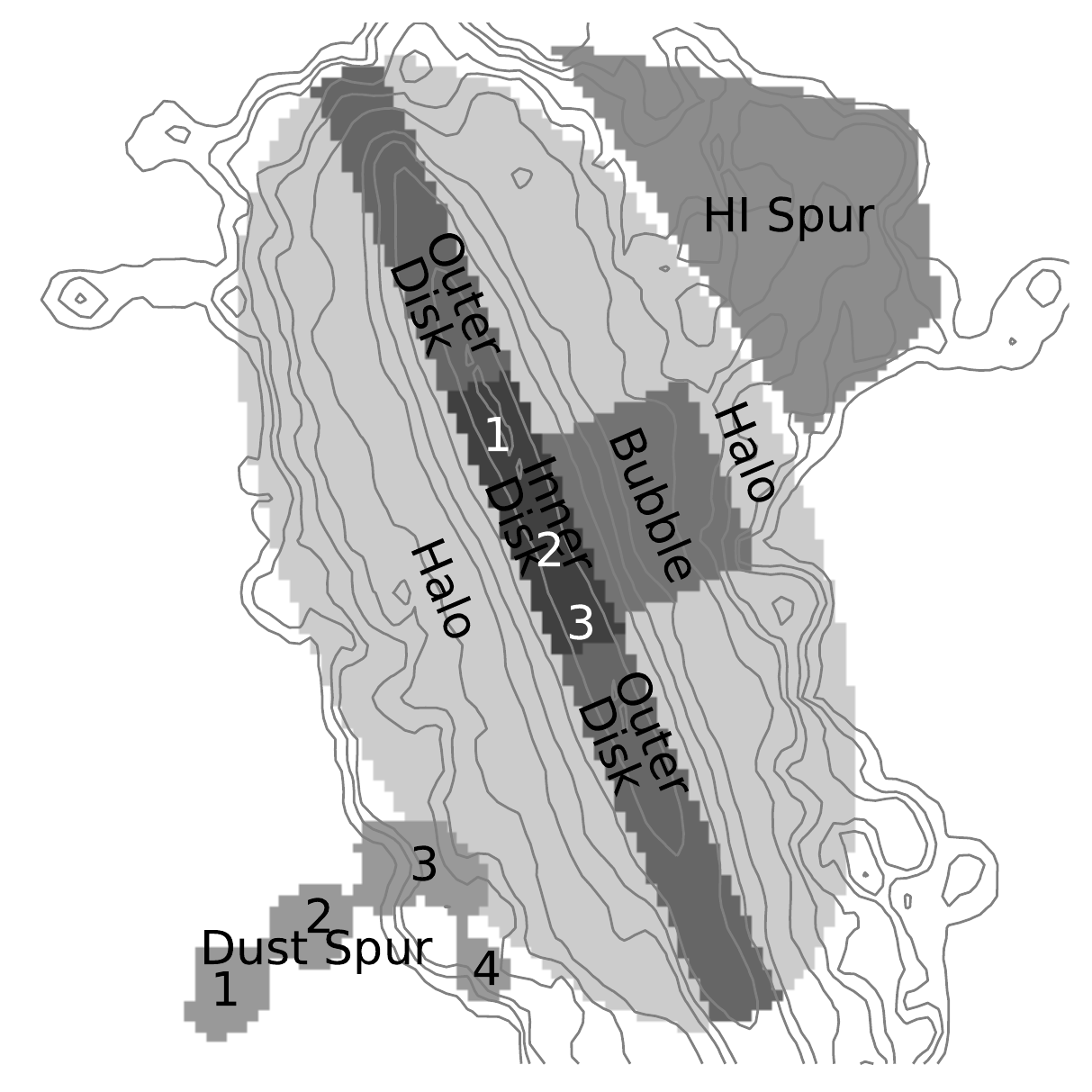}
\caption{Apertures used for far-infrared photometry shown
relative to \HI\ 21-cm contours from \citet{Oosterloo2007a}. The semi-major and semi-minor axises are 7\farcm4 (21 kpc) and 3\farcm7 (10.5 kpc), respectively.
We define the thickness of the disk by the thick stellar disk with the width of $\pm 0\farcm5$ (1.4 kpc). The inner disk has a width of 4\farcm1 (11.7 kpc) which includes the three star-forming clumps. The other rectangle marks the aperture we use for the superbubble photometry. The height and width of the superbubble are 2\farcm7 (7.7 kpc) and 2\farcm8 (8.0 kpc), respectively.
The ellipse encompasses the extended halo emission in the lowest 
resolution band. 
We define the aperture for the dust spur southeast of the galaxy 
by the 250\um\ contour at 1.7 mJy/pix, which is  3$\sigma$ of
the background noise, and outside the extended dust halo.
We define an aperture northwest of the 
galaxy in order to place upper limits on the far-infrared emission 
from the \HI\ spur.
Each dust clump and hot spot (star forming clump) listed in Table~\ref{tab:fir_radio} is labeled with a number.
}
\label{fig:region}
\end{center}
\end{figure}

In the next section, we validate our approach using the spectral energy distribution for
the entire galaxy. In Section~\ref{sec:sed_components}, we then describe the dust properties of the 
morphological  components identified in the previous section and defined in Figure \ref{fig:region}. We begin with the dust mass and temperature
of each region as a whole, and then we explore these properties on a pixel-by-pixel basis.
We estimate dust-to-gas ratios for the various components in Section~\ref{sec:d2g}.

\subsection{Validation of the Spectral Energy Distribution Fitting} 


The top panel of Figure~\ref{fig:sed_fit} shows the SED for the entire galaxy.
We fit only the 70 $-$ 500\um\ data because the 24\um\ flux comes from a 
warmer dust component. 
The single-temperature, MBB model with $\beta \equiv 2$
gives a dust mass of $M_{d} = 1.1 \times 10^8$\msun\ and dust
temperature $T =21.7\pm0.2$~K.  We also plot the fit with $\beta$
taken as a free parameter. A comparison shows that our fitted dust
parameters are robust to minor changes in the dust emissivity; see
Table~\ref{tab:dust} for a quantitative comparison.

The dominant error term is the 10\% uncertainty in the flux calibration.
To determine the uncertainty on the fitted mass and temperture, we 
scale the photometry by 10\% and refit the models. We find that
systematically increasing (or decreasing) all the fluxes can increase (decrease)
the fitted dust masses by $\sim$10-20\%, but dust temperature varies by less than a degree.
Our results confirm  the dust mass and temperature obtained previously by \citet{Hughes2014,Bocchio2016}.


The total dust mass fit to the FIR observations is nearly twice
that inferred from radiative transfer modeling of the distribution of 
optical and near-IR emission. For example, when scaled to our adopted distance, 
\citet{Xilouris1998} found $5.3 \times 10^7$\msun\ of dust. A significant fraction
of the dust presumably goes undetected because little optical/near-IR radiation emerges
from some regions of the galaxy. Previous work has suggested
that this hidden dust could reside in a  geometrically thinner dust disk \citep{Popescu2000,Popescu2011} or 
in clumpy clouds associated with molecular gas \citep{Bianchi2008}.
\citet{Seon2014a} considered \n891 specifically and introduced a thin dust disk of
scale height $\approx 5\arcsec$ to simultaneously describe the distribution of GALEX UV 
emission \citep{Morrissey2007}  and {\it Herschel} SPIRE photometry \citep{Bianchi2011}.


For completeness, we point out that
the impact of smaller dust grains (grain radii $< 50$ \AA) on our fits
is constrained by the 24\um\ flux. We fit a two temperature model to the 
full SED including the 24\um\ flux.  The temperature of a warmer component is not 
well constrained by this single data point, so we assumed a dust temperature of 
$T=220$~K to facilitate comparison to
\n4631. In Paper I, we found an additional dust component at 220 K in \n4631. 
In their single-temperature, MBB fits, they treated the 70\um~flux as 
an upper limit because the contribution from the warmer component 
was non-negligible. We find, however, that the contribution of
this warmer component to the 70\um~flux is only 0.3\% for this 
two temperature model fitted to \n891. 
Assuming that the peak wavelength of the MBB curve for the warmer component is shorter than 24\um, we vary the temperature of the warmer component, and the contribution to 70\um\ flux is 1.3\% at most. Therefore, we consider 70\um\ data point in the same way as other wavelength data in the fitting process.
For an assumed warm dust temperature
of 220~K, we fit a two-component model and find dust mass in the 220 K component is negligible ($\sim0.00003$\%) compared with dust mass in the $\sim22$ K component. The mass in the cool component in this two-component fitting does not deviate from the mass in the one-component, cool component only, fitting as shown in Table~\ref{tab:dust}.



\subsection{SED Fitting of Dust Components} \label{sec:sed_components}

\begin{figure} 
 \begin{center}
    \includegraphics[width=0.5\columnwidth, angle=0, trim = 100 0 100 0]{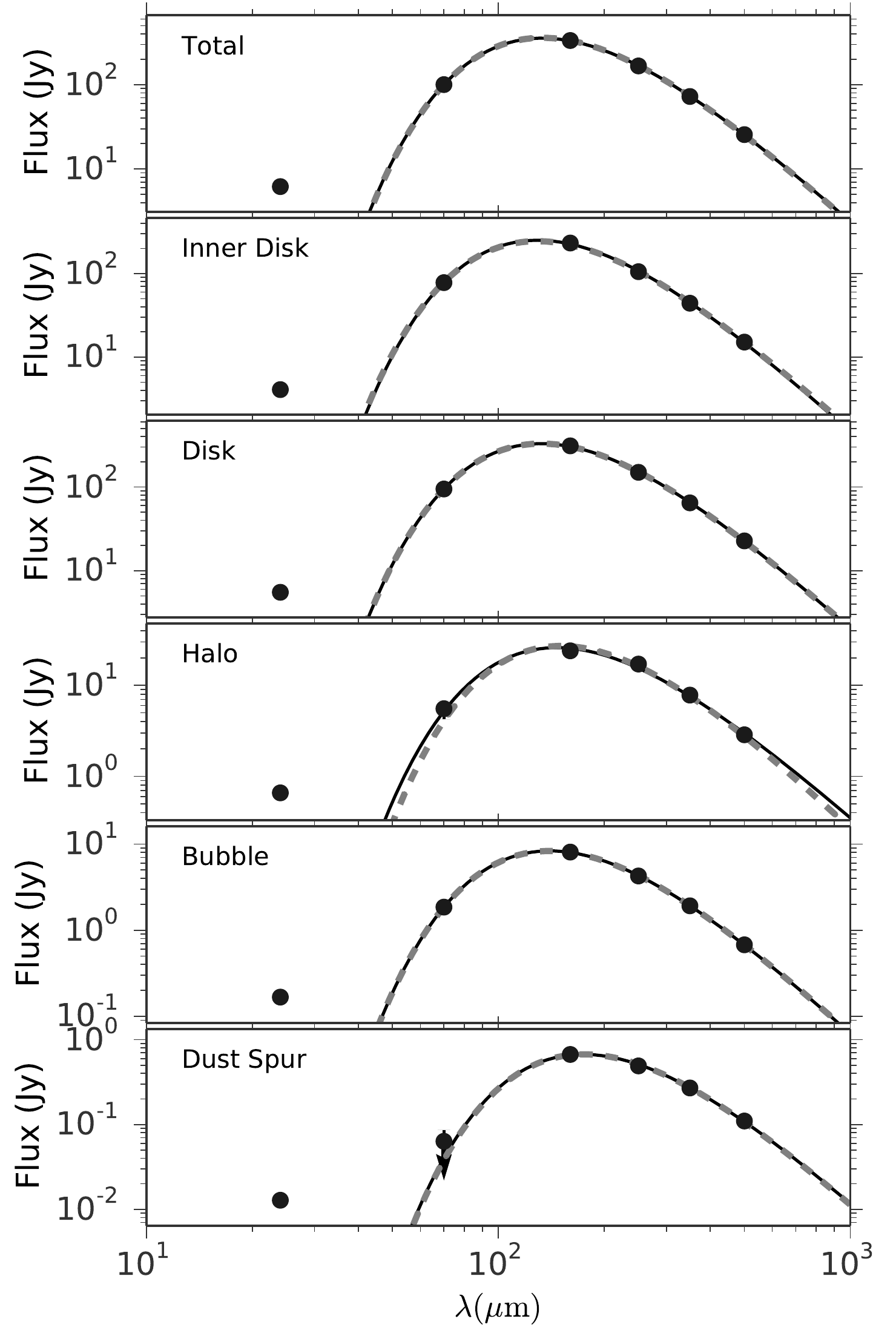}
   \caption{The best fit MBB model for integrated flux of each part of NGC 891 in 
24\um, 70\um, 160\um, 250\um, 350\um, and 500\um\ bands. 
The regions are defined in Figure \ref{fig:region}. The flux for 24\um\ is not used for the fitting as it traces different dust components. Gray-dashed line is the MBB model fits when $\beta$ is fixed at 2.}
    \label{fig:sed_fit}
     \end{center}
      \end{figure}

Table~\ref{tab:dust} summarizes the dust properties for the regions labeled in Figure~\ref{fig:region}.
The photometry shown in Figure~\ref{fig:sed_fit} was obtained after 
degrading the resolution of the 70, 160, 250, and 350\um\ maps to match that of the 500\um\ map; 
see details in  Section~\ref{sec:observations}. 


Our analysis constrains the contribution from dust well beyond the galactic plane where 
the starlight is too faint to model its attenuation. 
Inspection of  Table~\ref{tab:dust} indicates 87\% of the dust mass is associated with
the galactic disk (\logMdust $=7.970\pm0.006$) with 65\% of the disk dust 
concentrated in the inner disk (\logMdust $=7.781\pm0.012$). 
We associate roughly 13\% of the total dust mass with the halo component at
$|z|>0\farcm5 $, or $\pm\ 1.42$ kpc; the same halo aperture includes 20\%
of the total \HI\ mass. This difference can be attributed to the lower dust-to-gas ratio
of the halo.  Roughly 25\% of the halo dust mass (\logMdust $=7.163\pm0.053$) comes from
the superbubble region, while the dust spur, \logMdust $= 5.88\pm0.02$, makes a smaller
contribution.

The halo dust temperature ($T=19.24\pm0.62$~K) is significantly cooler
than the disk dust ($T=21.89\pm0.06$~K). The superbubble dust ($T=20.61\pm0.03$~K)
is warmer than the surrounding halo but cooler than the disk.
The emission from the dust spur is weak at 70\um\ compared to the SPIRE bands, and 
the fitted dust temperature is considerably lower than that of the disk or halo.


%
%


\begin{table*}
	\centering
	\caption{Dust properties derived from the MBB model fitting.}
	\label{tab:dust}
	\begin{tabular}{ccccccccc} %
		\hline
Component$^{d}$ & Total & Inner Disk & Disk & Halo & Superbubble & Dust Spur & \HI\ Spur \\
		\hline
log ${\rm M}_{\rm dust}$ & 8.025$\pm$0.005 & 7.781$\pm$0.012 & 7.970$\pm$0.006 & 7.163$\pm$0.053 & 6.468$\pm$0.003 & 5.883$\pm$0.019 & $<3.443$\,$^{a}$ \\
$\prime\prime$ & 8.018$\pm$0.007 & 7.800$\pm$0.018 & 7.970$\pm$0.013 & 7.061$\pm$0.074 & 6.474$\pm$0.004 & 5.868$\pm$0.061 & \ldots \\
$\beta$ & 2 & 2 & 2 & 2 & 2 & 2 & 2\\
$\prime\prime$ & 1.97$\pm$0.03 & 2.09$\pm$0.07 & 2.00$\pm$0.05 & 1.66$\pm$0.22 & 2.03$\pm$0.02 & 1.96$\pm$0.16 & \ldots\\
T (K) & 21.72$\pm$0.05 & 22.49$\pm$0.12 & 21.89$\pm$0.06 & 19.25$\pm$0.62 & 20.61$\pm$0.03 & 16.57$\pm$0.17 & 19.25$^{a}$ \\
$\prime\prime$ & 21.89$\pm$0.15 & 22.02$\pm$0.37 & 21.89$\pm$0.26 & 21.30$\pm$1.45 & 20.50$\pm$0.08 & 16.79$\pm$0.81 & \ldots \\
$\chi_{\nu}^2$ & 0.02 & 0.11 & 0.03 & 1.05 & 0.01 & 0.08 & \ldots \\
$\prime\prime$ & 0.02 & 0.09 & 0.05 & 0.75 & 0.01 & 0.11 & \ldots \\
log $({\rm M}_{\rm HI}/{\rm M}_{\odot}$)\,$^{b}$     & 9.64 & 9.09 & 9.48 & 9.10 & 8.07 & $<7.22$ & 7.20 \\
${\rm M}_{\rm dust}$/${\rm M}_{\rm gas}$\,$^{c}$ & 0.0088 & 0.0082 & 0.0095 & 0.0084 & 0.025 & $>0.0046$ &  $<0.00017$ \\
$\prime\prime$                       & 0.0087 & 0.0086 & 0.0095 & 0.0067 & 0.025 & $>0.0044$ &  \ldots \\
		\hline
	\end{tabular}
\\
$^{a}$ 
The dust mass limit is computed by rescaling the MBB
  model to satisfy the FIR flux limits of the
 \HI\ 
spur assuming
  $\beta=2$ and the halo dust temperature.
$^{b}$ {The \HI\ masses were  measured from \HI\ map of \citet{Oosterloo2007a}.}
$^{c}$ {The gas mass was computed as $M_{\rm gas} = 1.36 (M(\HI) +
  M({\rm H_2}))$, where the coefficient accounts for the gas mass in
  helium. }
$^{d}$ {Alternate rows compare fixed and free spectral indices for the dust opacity.}
\end{table*}


\subsection{Pixel by Pixel Analysis} \label{sec:pixel_by_pixel}

\begin{figure}
\begin{center}
  \includegraphics[width=\linewidth]{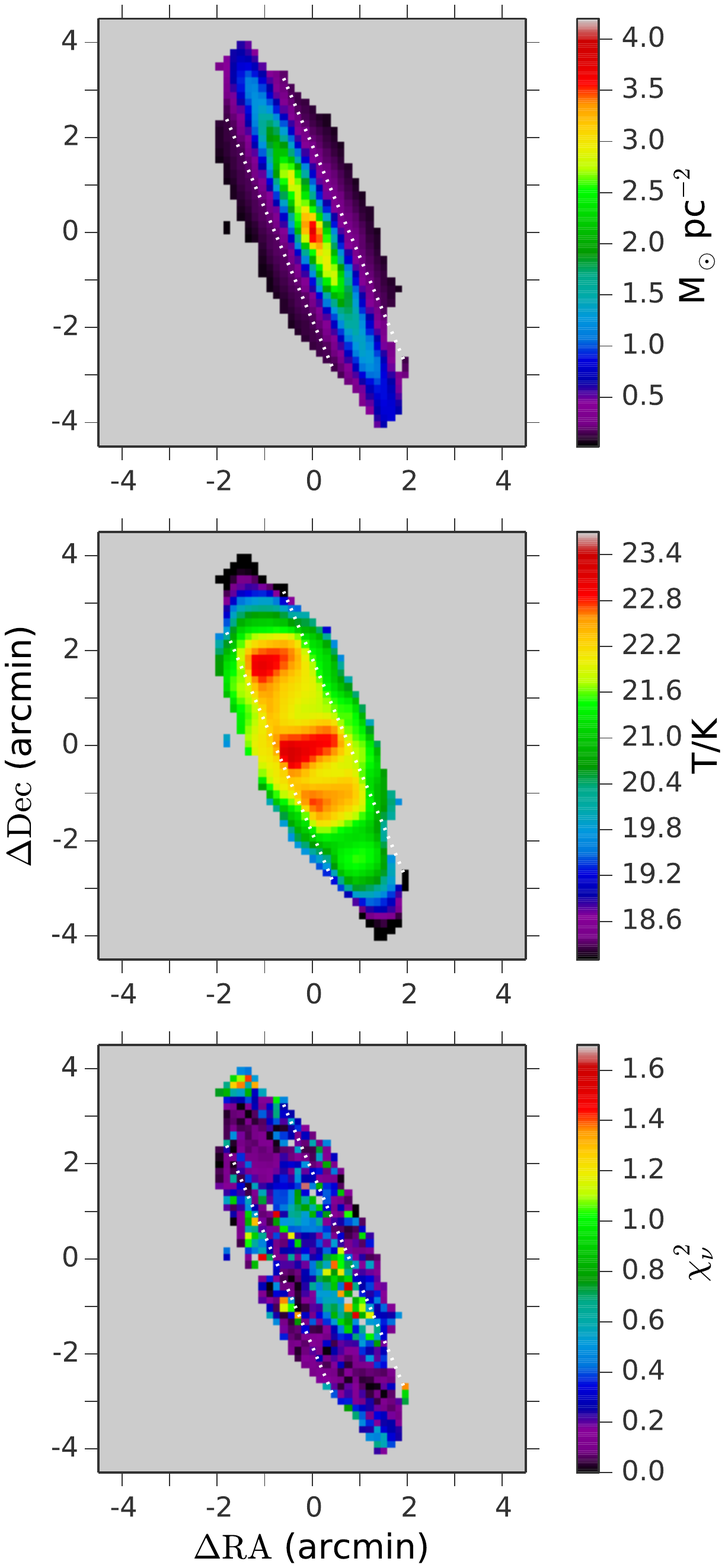}
\caption{Pixel-to-pixel dust mass, dust temperature, and
$\chi_{\nu}^2$ of the model fit with $\beta \equiv 2$ 
are shown. Each pixel is $9\farcs0
\times 9\farcs0$. The disk boundary defined in Figure~\ref{fig:region} 
is indicated by the white-dotted line in each panel. 
}
\label{fig:pix2pix}
\end{center}
\end{figure}

We also examined the variation in dust properties at the resolution limit of
the 500\um\ image. We fit the SED of all pixels that had  $S/N > 3\sigma$ in all five bands, 
essentially the entire disk component and the disk -- halo interface.
\citet{Hughes2014} previously found the variation in dust temperature to be more uneven
than that of the dust mass. Figure~\ref{fig:pix2pix} shows our resulting maps of the temperature
and dust surface density, i.e., the column density of dust in mass units.

The dust surface density decreases with distance from the nucleus along the major axis.
In the vertical direction,  the dust surface density
drops very quickly to values less than 0.50\msun\ pc$^{-2}$.

\begin{figure*}
\begin{center}
  \includegraphics[trim={0 0 0 0}]{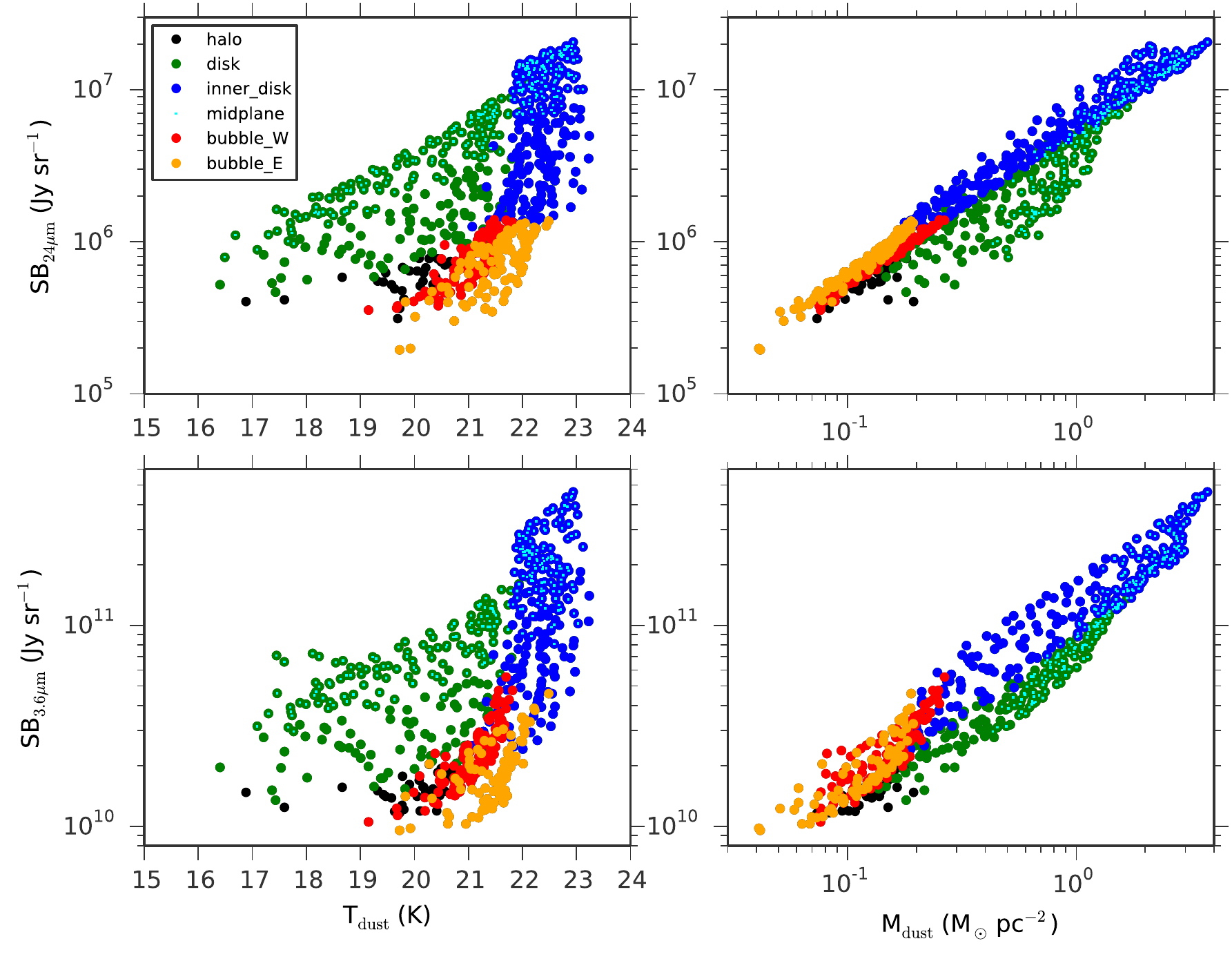}
\caption{
Pixel-to-pixel correlations between dust properties and shorter wavelength
emission not used in the SED fitting. The regions shown correspond to the map
in Figure~\ref{fig:pix2pix}. In this plot, {\it midplane} is defined as the
region within $|z| \le 0\farcm7$ (1.99 kpc) of the disk plane.
}
\label{fig:dust_property}
\end{center}
\end{figure*}

The temperature range in Fig.~\ref{fig:dust_property} 
is similar to the $T_d = 17-24$~K range found previously by \citet{Hughes2014}. 
The hotspots identified in both the
24\um\ map and the 70\um/160\um~color map  stand out as regions with 
$T_d \ge\ 22.5$~K in Figure~\ref{fig:pix2pix}. This region of the disk is
surrounded by a molecular ring \citep{Israel1999a,Scoville1993a}.
The dust cools off with increasing radius in the disk beyond the molecular ring.

We compare the dust temperature and surface density directly to the 3.6\um\ and 24\um\ emission in 
Figure~\ref{fig:dust_property}.  The 24\um\ emission is a good proxy for
the star formation rate surface density \citep{Calzetti2007}, while the
3.6\um\ emission traces the stellar mass density. We have color coded the 
points in Figure~\ref{fig:dust_property} by their location in \n891.

The dust in the midplane of the inner disk has temperature, $T_D \approx 22-23$~K and a high 
mass column, as indicated by its location in  the upper right of Figure~\ref{fig:dust_property}. 
With increasing vertical height at fixed
radius, the temperaure decreases by roughly one degree as the surface brightness drops by an order of magnitude. 
This locus (blue points) defines a narrow band that continues smoothly into the emission from the superbubble 
region where the temperature drops a few more degrees (orange and yellow points).  \citet{Hughes2014} produced
color maps that identify this same spectral transition with height above the disk. Adding the halo emission
immediately above the inner disk extends this correlation to dust temperatures $T_D \sles 20$~K. Based on
this correlation between dust temperature and 24\um\ emission, the dust in and above the inner
disk is heated primarily by the starlight from the star-forming regions. 

The width of the  SB(24\um) -- $T_D$ locus in the inner disk correlates with radial distance from
the nucleus.  For example, following the midplane points in Figure~\ref{fig:dust_property} from
the inner disk to the outer disk (green points) follows the upper edge of the triangular locus. At
fixed 24\um\ surface brightness, dust in the  midplane of the outer disk is cooler than dust above
the plane at small radii. Equivalently, at fixed dust temperature, the superbubble and halo emission
are not as bright at 24\um\ as the outer disk.  We expect a transition in the outer disk toward
heating by an older stellar population, and it seems plausible this transition manifests as the broad
triangle of green points in  Figure~\ref{fig:dust_property}. Consistent with this interpretation, the
3.6\um\ emission from the midplane points in the outer disk (green points with cyan dot) are strongly
correlated with the dust temperature. Dust above the midplane of the outer disk fills the interior
of the triangular locus, as might be expected for dust heated by a mixture of the two stellar populations.

We acknowledge that the temperature correlations look nearly identical against
the near-infrared (3.6\um) and mid-infrared (24\um) emission, so the temperatures have the 
same correlations with the tracers of stellar mass and star formation, respectively. It seems likely
that this happens because the mass and SFR both decrease with separation from the center of
the galaxy.

The panels on the right side of Figure~\ref{fig:dust_property} show  power law
correlations between infrared emission and dust surface density. For the inner disk, 
the correlation is tighter in the mid-infrared (24\um), more closely related to SFR surface density.
In the outer disk, the 3.6\um\ emission shows the stronger correlation, consistent with a closer relation
to stellar mass surface density.

\subsection{Dust-to-gas Ratio} \label{sec:d2g}


Table~\ref{tab:dust} compares the dust mass to the gas mass, where all 
measurements taken from the literature have been scaled to  an \n891  
distance of 9.77~Mpc. The total \HI\ mass of \n891 is then $4.4 
\times 10^9$\msun \citep{Oosterloo2007a}. The total CO(1-0) intensity 
\citep{Scoville1993a,GarciaBurillo1992} correspondes to a molecular gas mass
of  $M({\rm H_2}) \approx 4.2 \times 10^9\msun$ for the \citet{Bolatto2013}
calibration of the CO-to-$\rm H_2$ conversion factor, 
$X_{\rm CO} =  2 \times 10^{20} {\rm cm}^{-2} {\rm [K~km~s^{-1}]}^{-1}$. 
The ratio of molecular-to-atomic gas is close to unity, $M({\rm H_2})/M(\HI) \approx 0.95$, 
which is higher than the median of 0.6 found for Sab galaxies  \citep{Obreschkow2009a}.  
Including the mass contribution from helium, the total mass of cold gas
is $1.2 \times 10^{10}$\msun.

Comparing our measured dust mass to the atomic and molecular gas mass, we find a global gas-to-dust ratio
$M_{\rm gas}/M_{\rm dust} \approx 110$,
where the uncertainty introduced by the $X_{\rm CO}$-factor is 30\% \citep{Bolatto2013}.
The gas-to-dust ratios for the SINGS galaxies range from 100 to 340 after scaling the values in
Table~4 of \citet{Draine2007a}  to the same $X_{\rm CO}$ and including the helium mass. On this
scale, the gas-to-dust ratio of the Milky Way is 200 based on 
the $M_{\rm dust}/M_{\rm H} \approx 0.007$ ratio given by \citet{Draine2007a}.
Based on its atomic and molecular gas content, \n891 is dustier than the median Sab galaxy.

The molecular gas is concentrated in the inner disk of \n891, so it is not surprising that 
the inner disk is the dustiest region. The \HI\ disk is larger than the molecular disk.  
\citet{Oosterloo2007a} attribute slightly over 70\% of the total \HI\ to the disk, and 
our measurements of their \HI\ map indicate the inner disk \HI\ mass is $1.22 \times 10^9$\msun,
slightly less than half the mass of the entire \HI\ disk. The total gas mass of the inner (and
full) disk  is then $7.4 \times 10^9$\msun\ ($9.8 \times 10^9$\msun). These disk masses do not
include the small contribution from warm ionized gas \citep{Rand1990,Dettmar1990}, but 
they do include the mass contribution from helium, as described in the notes to 
Table~\ref{tab:dust}.

The filament northwest of \n891 contains at least $1.6 \times 10^7$\msun\ of  \HI
\citep{Oosterloo2007a}. No thermal dust emission is detected from the \HI\ filament.
The upper limit on the dust mass excludes gas-to-dust ratios similar to galactic
disks.  If the \HI\ filament is accreting gas as \citet{Oosterloo2007a} suggest, then
the lack of dust emission favors low metallicity gas.  Alternatively, if the \HI\
filament is fountain gas, the timescale to recycle the disk gas must be long enough
to destroy even large grains, which seems unlikely based on the large amount of dust
that persists in the CGM \citep{Menard2012a}.

The properties of the dust spur southeast of \n891 contrast sharply with
those of the \HI\ filament.  At the distance of \n891, the far-infrared filament contains nearly 
$8 \times 10^5$\msun\ of dust. Based on the dust mass, this circumgalactic gas was once in a
galaxy, either \n891 or a satellite.  Using the \HI\ map \citep{Oosterloo2007a}, 
the \HI\ mass in the dust spur is less than $2 \times 10^7$\msun.  The exceptionally low 
ratio of neutral hydrogen gas to dust is a puzzle. Perhaps a background source will be
found for the infrared emission.  An alternative, however, is that most of the gas mass is
either molecular or ionized.

%



The mass ratio of \HI\ to dust in the inner CGM of \n891 is indistinguishable from the 
galactic disk.  We find this surprising because the neutral fraction of halo gas is 
around 1\% \citep{Popping2009,Bland-Hawthorn2017}.  The dust was presumably made inside galaxies
and transported into the halo via outflows or stripping.  Gas clouds are normally
destroyed by these processes unless special conditions are met, so we would not expect
the galactic gas to remain neutral. The inner CGM of \n891 contains more neutral 
gas than the median COS (Cosmic Origins Spectrograph) Halos galaxy \citep{Prochaska2017,Das2020}.  The {\it disk-like}
ratio of \HI\ to dust certainly suggests that the excess of neutral gas in the inner
CGM of \n891 was removed from galaxies. We will discuss the extrapolated dust mass of the CGM
in Section~\ref{sec:extrapolate_halo}.


\begin{table}
	\centering
	\caption{FIR and Radio Fluxes within \n891}
	\label{tab:fir_radio} 
	\begin{tabular}{lccccc} %
		\hline
Part &
log $L_{\rm FIR}$ &
log $L_{\rm 60\mu m}$ &
log $L_{\rm Cont}$ &
$q_{\rm IR}$ &
$q_{\rm 60\mu m}$ \\
&
(\lsun) &
(\lsun) &
($\rm W~Hz^{-1}$) &
&
\\
(1) &
(2) &
(3) &
(4) &
(5) &
(6)\\
		\hline
Total & 10.44 & 9.64 & 22.00 & 2.45 & 2.19 \\
Inner Disk & 10.30 & 9.55 & 21.75 & 2.56 & 2.32 \\
Disk & 10.41 & 9.62 & 21.87 & 2.55 & 2.29 \\
Halo & 9.28 & 8.32 & 21.39 & 1.90 & 1.56 \\
Bubble & 8.75 & 7.82 & 20.81 & 1.96 & 1.64 \\
Dust Spur & 7.60 & 6.04 & 19.13 & 2.48 & 1.93 \\
Dust Spur1 & 7.23 & 6.08 & 18.68 & 2.56 & 2.13 \\
Dust Spur2 & 6.91 & 5.27 & 17.68 & 3.24 & 2.71 \\
Dust Spur3 & 7.12 & 5.50 & 18.78 & 2.35 & 1.76 \\
Dust Spur4 & 6.56 & 5.18 & 18.30 & 2.27 & 1.76 \\
Hotspot1 & 9.76 & 9.02 & 21.20 & 2.57 & 2.33 \\
Hotspot2 & 9.80 & 9.08 & 21.24 & 2.57 & 2.34 \\
Hotspot3 & 9.56 & 8.80 & 20.98 & 2.59 & 2.34\\
		\hline
	\end{tabular}
\\
Columns: (1): Part of NGC 891, (2): IR luminosity at
  8\um\ - 1000\um\ from SED fit, (3): 60\um\ flux computed from the
  SED fit, (4): radio continuum flux from the continuum map by
  \citet{Oosterloo2007a}, (5) $q_{\rm IR} = {\rm log} \left( {\rm
    IR}\over{3.75\times10^{12}~{\rm W~m^{-2}}} \right) - {\rm log}
  \left( {S_{\rm 1.4~GHz}\over{\rm W~m^{-2}~Hz^{-1}}} \right) $ where
  IR is the FIR flux and $S_{\rm 1.4~GHz}$ is flux density at 1.4 GHz.
  (6): For $q_{\rm 60\mu m}$, the flux at 60\um\ is used. This
  definition is adopted from \citet{Condon1991a,Yun2001a,Bell2003a}.
\end{table}

\section{Discussion} \label{sec:discuss}

\subsection{The Dusty Superbubble}
\label{superbubble_wind.sec}

We found an extraplanar, dusty region that is significantly warmer than the
dusty halo. We called it the dusty superbubble because its location is coincident
with thermal X-ray emission ($kT = 0.71$~keV) likely associated with an outflow.
In Section~\ref{sec:blowout} below, we show that the thermal energy of this
superbubble is sufficient for the superbubble to punch through the disk and
blowout into the halo. Blowout in \n891 is of particular interest because the
SFR surface density is a factor of three below the commonly assumed threshold
of 0.1 \msunyr kpc$^{-2}$. The fate of this outflow depends largely on the
halo gas density profile, which is reasonably well constrained in \n891. We present
this evolution in Section~\ref{sec:wind_fate} and discuss its relationship to
the recent detection of cosmic ray advection in the halo of \n891.

\subsubsection{A Galactic Wind at SFR Surface Density $< 0.1$ \msunyr kpc$^{-2}$}
\label{sec:blowout}


The first numerical simulations of superbubble blowout predicted that 
a superbubble would grow to heights of one to two vertical scale heights before the extraplanar shell accelerated and
broke up \citep{MacLow1988,MacLow1989}. 
Given the scale height measurements for \n891 \citep{Bocchio2016},
the height of the superbubble (2\farcm7) is 8 times the scale height of the dust/gas thick disk 
(0\farcm34 in 24\um) and 68 times that of the dust/gas thin disk (0\farcm04 in 24\um).

We will therefore discuss whether the \n891 superbubble will develop into a wind. The star formation rate surface 
density is elevated in the disk of \n891, similar to that in a typical disk galaxy several Gyr ago. Whereas the galaxy M82
has become the prototypical starburst outflow, we suggest that the galaxy \n891 provides a  more typical example
of star-formation feedback in well-developed galactic disks.
To better understand how winds are launched, sophisticated simulations that include  multiphase gas, cosmic
rays, and turbulence should be tested against the observed properties of \n891 as well as starburst galaxies.

%

The pioneering work of \citet{MacLow1988,MacLow1989} predicts that when a superbubble grows
to scales comparable to the gas scaleheight, the expansion velocity $V(r)$ of the shell  determines
its fate.  Shells moving faster than the sound speed of the ambient medium will accelerate
as they break through the disk; hydrodynamic instabilities then disrupt the shell, and a
free-flowing galactic wind develops \citep{CC85}. Shells with $V(H) \le\ c_s$ will continue
to decelerate and will eventually collapse due to disruption caused by differential rotation
and ISM turbulence.  The critical rate of energy injection for blowout to occur depends
on the vertical gas density distribution.

Eqn.~5 of \citet{Strickland2004b} gives the critical power for an exponential gas distribution,
$\rho(z) = \rho_0 \exp(-z/H)$, in terms of the gas scale height $H$. The scale height of 
the thin, molecular disk in \n891 is just 83~pc \citep{Scoville1993a}. 
The \HI\ disk has a thickness $H \approx 2.6$~kpc \citep{Oosterloo2007a}, and 
warm ionized gas extends 4.5 kpc above the disk plane \citep{Dettmar1990}.
Scaling to the thick \HI\ disk, the critical mechanical
power increases to 
\begin{eqnarray}
L_{crit} = 2.3 \times 10^{39} {\rm ~ergs~s}^{-1} 
\left( {1 {\rm ~cm}^{-3}}\over{n_0}   \right)^{1/2} \times \\ \nonumber
\left( {P_0/k}\over{10^4 {\rm ~K~cm}^{-3}}  \right)^{3/2} 
\left( {H}\over{2.6 {\rm ~kpc}}  \right)^2.
\end{eqnarray}


The rate of mechanical power injection from supernovae and stellar winds can be
estimated from the SFR. We adopt the solar metallicity models from Starburst~99
and the Salpeter initial mass function. For a star formation history with a 
constant SFR,  the steady-state feedback power is $L_{\rm w} = 10^{41.8}$~ergs~s$^{-1}$
for a SFR of 1\msunyr \citep{Leitherer1999}. This calibration includes only 
stars in the mass range $1 \le\ {\rm M_*}/\msun \le\ 100$. To be consistent with 
the mass range of $0.1 \le\ {\rm M_*}/\msun \le\ 100$ used to describe the SFR 
\citep{Kennicutt1998a}, we divide by a factor of 2.55, obtaining 
$L_{\rm w} = 10^{41.4}$~ergs~s$^{-1}$ for a SFR of 1\msunyr. 

We have listed far-infrared luminosities for the entire galaxy and the inner disk
in Table~\ref{tab:fir_radio}.  Applying $SFR (\msunyr) = 4.5  \times 10^{-44} L_{\rm FIR} 
({\rm ergs~s}^{-1}~{\rm Hz}^{-1})$ \citep{Kennicutt1998a}, the rate of obscured star formation 
for the entire galaxy is 4.8\msunyr. This is technically a lower limit on the SFR because
not all of the UV light from \n891 is obscured by dust. However, we obtained the FUV and NUV magnitudes 
of \n891 from the GALEX nearby galaxy catalog and found that the contribution from unobscured SFR is a
relatively small, roughly $0.2$\msunyr.

The mechanical energy produced by massive stars in the inner disk is most relevant 
to the evolution of the superbubble. We measure an inner disk SFR of 3.5\msunyr\ from 
the 24\um\ emission. The available mechanical power is then $L_{\rm w} = 8.7 \times 10^{41}$~ergs~s$^{-1}$,
and $L_{\rm w} / L_{\rm crit} \approx 380$.  Even a few percent of the mechanical 
energy produced in the inner disk is sufficient to power the expansion of a superbubble that
will blowout of the disk of \n891.


Theoretical studies have tried to predict the critical SFR surface density 
for the formation of a galactic wind.  Overcoming radiative losses has been
a major hurdle when star-forming regions are embedded in realistic gas clouds.
\citet{Murray2011}  considered the role of radiation pressure from massive star 
clusters in launching winds and suggested 0.1\msunyr\ per kpc$^2$ as a threshold. \citet{Scannapieco2013} and \citet{Hayward2017} both connected blowout directly to the turbulence in the ISM.

Empirically, \citet{Heckman2002} suggested 0.1\msunyr\ per kpc$^2$, based largely on 
nearby starburst galaxies. When they were younger, however, essentially all disk
galaxies drove massive winds \citep{Martin2012,Rubin2014}. \citet{Kornei2012}
measured SFR surface densities 
for redshift $z\sim 1$ star-forming galaxies. Roughly  30\% of the outflow galaxies 
has SFR surface densities below the suggested threshold of 0.1\msunyr\ per kpc$^2$.
We suggest that \n891 is an excellent nearby analog, where a
wind is developing at a 
SFR surface density (SFRSD)
a factor of a few lower than the threshold 
suggested by \citet{Heckman2002}. Normalizing the 24\um\ luminosity of the inner 
disk by the area of a disk of diameter 4\farcm1, we obtain a SFRSD within the molecular 
ring of $\Sigma = 0.031$\msunyr~kpc$^{-2}$ for \n891.
In comparison, for NGC 4631 (Paper I), the SFRSD is estimated to be 
$\Sigma = 0.15$\msunyr~kpc$^{-2}$ as the SFR is 2.9\msunyr\ within a radius of 2.5 kpc 
where almost all of the star formation is taking place, or a SFR surface density
about 5 times higher than that for NGC 891.

\citet{Hayward2017} describe the condition for blowout by a critical gas fraction, 
which they find is  30\%. We showed in Section~\ref{sec:d2g} that the total gas mass 
of the inner disk is $7.4 \times 10^9$\msun. We now place bounds on the stellar mass.
The  IRAC 3.6\um\ flux of the inner disk requires $M_* \sgreat 4 \times 10^{10}$ \msun, 
a lower limit due to optical depth effects. The rotation speed requires a dynamical 
mass $M_{\rm Dyn}  \approx 6.9 \times 10^{10}$\msun\ within a radius of $R=5.83$~kpc. 
The gas fraction in the inner disk of \n891 is in the 10-16\% range, a factor of two
below the \citet{Hayward2017} threshold.



We found that the dusty superbubble is spatially coincident with hot gas ($kT = 0.71 \pm 0.04$~kev) that
is concentrated above the star-forming regions in \n891 \citep{Hodges-Kluck2018}. This soft X-ray emission
very likely comes from the interaction of a hotter wind with cold gas. The cold gas mass is roughly $3 \times 10^8$\msun\
based on the outflowing dust mass,   $3 \times 10^6$\msun, and the disk gas-to-dust ratio.  Hot winds have very
low emission measure and have only been directly detected in a couple of starburst galaxies \citep{Strickland2004a,Strickland2004b}. It
is therefore of significant interest that the central star-forming region in \n891 is surrounded by
diffuse hard X-ray emission \citep{Hodges-Kluck2018}, possibly related to the hot wind. Regardless of the origin
of this hard component, however, \n891 shows all the signatures of a thermally driven wind.


\subsubsection{Interaction of Wind-Driven Bubble with CGM}
\label{sec:wind_fate}


As the thermal wind plows into the CGM, it does work on the halo gas and also cools. The classic wind model 
widely applied to starburst galaxies \citet{CC85} does not adequately capture this wind -- halo interaction
because it neglects gravity, radiative losses, and the density profile of the CGM. These processes 
ultimately determine whether the wind cools, whether a galactic fountain forms,  and whether 
outflowing gas escapes the gravitational potential.
Hydrodynamic simulations have not yet spatially resolved the production and evolution of cool gas
throughout a galactic disk and the surrounding CGM, but the wind -- halo interaction
has been explored over a broad parameter range, however, using semi-analytic models
\citep{Scannapieco2002,Furlanetto2003,Samui2008,Lochhaas2018}. 
We applied the \citet{Lochhaas2018} model to \n891 to gain further insight about the outflow.

The \citet{Lochhaas2018} model 
extends the classic structure of interstellar bubbles \citep{Weaver1977} to circumgalactic scales. 
The hot wind drives an expanding shock front that compresses the surrounding halo gas into a thin shell. 
A contact discontinuity separates this shell from the wind. The wind fluid is
described by the shocked wind immediately behind the shell, an interior region of unshocked
cool wind, and the innermost region of unshocked hot wind. The results suggest that
whether adiabatic cooling of the hot wind is accompanied by radiative losses depends on
the star formation rate, halo density profile, and wind density in a non-monotonic manner.

Figure 2 of \citet{Lochhaas2018} shows the evolution of the contact discontinuity and indicates
when the shocked wind begins to cool. The figure cannot be directly applied to NGC 891, however,
because the fiducial launch radius, $r_1 = 1$~kpc, corresponds to a SFRSD $\ge\ 0.3$ \msunyr ~kpc$^{-2}$,
considerably higher than the 0.031 \msunyr ~kpc$^{-2}$ in the central disk of NGC 891. Repeating the
calculation with $r_1 = 7$~kpc and SFR=4.5\msunyr\ shows that the NGC 891 outflow lies in a sweet spot
where the shocked wind cools radiatively; there is enough mass going into the shocked wind, yet the
expansion rate is not fast enough to dropping the density rapidly (C. Lochhaas, private communication).


%

After 20 Myr, the contact discontinuity has reached a height of 14 kpc and decelerated to $\approx 200$\kms.
The mean advection speed is therefore larger than the 150\kms fit to the spectral index profile of the 
halo radio emission \citep{Schmidt2019}. Whether this discrepancy is significant is unclear.
The semi-analytic model does not specify the physical mechanism launching the wind; it simply predicts how the
superbubble evolves for specified $\alpha$ and $\beta$.  The parameter $\alpha$ describes the efficiency at 
which supernova energy is transferred to the wind; the $\beta$ parameter describes the mass loading. Neither
parameter is well constrained, and both are assumed to be unity in the \citet{Lochhaas2018} calculations.
Values of $\alpha < 1$ would lower the launch velocity, however, so it is not obvious that small adjustments
could produce a better match between the velocities.



We have argued that the thermal pressure of the star-forming region is sufficient 
for blowout, but cosmic ray electrons may also be important for driving the \n891 outflow.  The 
non-thermal radio spectral index steepens nearly linearly with height just above the 
disk of \n891, a signature that the cosmic ray electrons are transported by advection 
rather than diffusion \citep{Mulcahy2018,Schmidt2019}. 
The non-thermal spectral index  flattens at heights above 2~kpc, consistent with 
adiabatic expansion dominating synchrotron losses across the region we call the superbubble.
\citet{Schmidt2019} argue that the cosmic ray advection plausibly reaches the 
halo escape velocity 9 to 17 kpc above the disk.


\subsection{The Origin of the Dust spur and the \HI\ Filament} \label{sec:explain_spur}

\subsubsection{The Dust Spur}

The {\it Herschel} imaging with PACS and SPIRE 
confirms the presence of a dust spur discovered in the {\it Spitzer}/MIPS 24\um\ image.
The shape of the far-infrared/sub-mm SED (see Figure~\ref{fig:sed_fit}) is consistent with emission from a 
structure associated with \n891.

The relative locations of the dust spur to the southeast and the \HI\ spur to the northwest is perplexing
in part because these two streams have distinctly different composition. The dust-to-gas ratio 
(see Table~\ref{tab:dust}) is high in the dust spur but very low in the \HI\ spur. The dust spur is
therefore likely composed of higher metallicity gas than the \HI\ spur. 

The dust spur contains roughly $7.6 \times 10^5$\msun\ of dust. If the dust-to-gas ratio is similar 
to the disk of \n891, then the gas mass is $\approx 9.5 \times 10^7$\msun. This mass exceeds the upper 
limit on the mass of neutral hydrogen, $M_{\HI} < 1.7 \times 10^7$\msun, in this region of the halo. 
This apparent contradiction may be explained by additional gas mass in another phase, a higher
dust-to-gas ratio in the spur (compared to the \n891 disk), or some combination of these properties.

Within the dust spur, the local maxima in the far-infrared maps coincide with several
 knots in the radio continuum map. From inspection of Figure~\ref{fig:spire_smooth}, we
defined four clumps (see Fig.~\ref{fig:region}) for which we provide photometry in Table~\ref{tab:fir_radio}. 
The radio and 60\um\ fluxes of these clumps lie on or just above the radio-IR relation defined by 
star-forming galaxies \citep{deJong1985,Condon1992,Yun2001a}. We therefore suggest that the dust 
spur contains young, possibly obscured, star clusters. Since radio and 60\um\ flux both 
underestimate SFR at low luminosities \citep{Bell2003a}, however, it is difficult to make an accurate
estimate of their SFR.

Could tidal forces exerted on the disk of NGC 891 by a satellite galaxy produce the dust spur?
We estimated the required size of a companion using the Dahari parameter \citep{Dahari1984} and
the projected separation between the tip of the dust spur and the center of NGC 891, 
$ S =7\farcm88$. The tidal force on NGC 891 scales as the the mass of the companion, $F_t \propto
M_c / R^3$, where $R$ is the distance between the galaxies, but the companion mass is unknown. 
\citet{Dahari1984} introduced a dimensionless parameter $Q \propto F_t$  which uses the diameter of 
the primary galaxy, $D_p = 14\farcm46$ 
\citep{rc3}, as a scaling parameter, obtaining 
\begin{equation}
F_t \propto Q \equiv\ \frac{(D_c D_p)^{1.5}}{ S^{3}},
\end{equation}
and empirically showing that a strong tidal interaction requires $Q \ge\ 1$. This argument
indicates that a satellite galaxy large enough to pull the dust spur out of the NGC 891 disk
would have a major axis $D_c \approx 4\farcm30$ or larger. This size is remarkably similar
to the extent of the dust spur, which is roughly 2\arcmin\ by 4\arcmin\ in Figure 2. Yet 
the old stellar population of this hypothetical satellite is not detected.

The dust spur may be the {\it smoking gun} of a dark satellite passing through the disk of 
\n891.
The Milky Way's population of high velocity clouds (HVCs) is thought to include a subpopulation that
is confined by dark matter minihalos \citep{Blitz1999}. The properties of compact, isolated HVCs are
similar in many respects to those of dwarf irreglular galaxies, but they lack a high surface brightness
stellar population \citep{Braun2000}. They have \HI\ masses of $10^5 - 10^6$\msun, i.e., well below the
upper limits on the \HI\ mass of the dust spur, but total masses of $10^7 - 10^8$\msun\ which are as large
as the masses of some dwarf galaxies \citep{Adams2013}. Hydrodynamical simulations of a minihalo
colliding with a disk show that the interaction pulls a coherent gas cloud out of the far side of the 
disk \citep{Nichols2014,Galyardt2016,Tepper2018}, 
a qualitatively different result than the collision of a pure
baryonic cloud with a disk.  
These structures persist for $\sim 60$~Myr \citep{Galyardt2016} or perhaps even longer 
\citep{Tepper2018}, trigger
star-formation in the minihalo, but eventually deposit the majority of the HVC's gas mass to the 
disk of the primary galaxy.


\subsubsection{The H I Filament}
\citet{Oosterloo2007a} considered several interpretations of the \HI\ spur. They considered gas
previously ejected from the disk and now returning to the disk in a galactic fountain and found 
the recycled gas had to lose significant angular momentum, perhaps to interaction from
infall from the intergalactic medium (IGM), to be consistent with the \HI\ kinematics. Alternatively,
they suggested the stream could be cold gas directly accreted from the IGM or condensing out of a hot, virialized
halo. No star formation has been associated with the \HI\ spur.

In our opinion, the substantial mass of the \HI\ filament,  $M(\HI) \approx 1.6 \times 10^7$\msun, 
suggests a discrete event. We suggest the  ram pressure from the gas halo in \n891 pushed the
gas out of a high-velocity cloud or  satellite galaxy as it approached the disk.
For purposes of illustration, we have estimated the halo density $n_0 \approx 4 \times 10^{-4}$~cm$^{-3}$
at $z = 12.5$~kpc and $R= 22.7$~kpc from the fitted \HI\ surface density model \citep{Oosterloo2007a}.
Models connecting the pressure to the
ratio of molecular-to-atomic hydrogen, for example, suggest $P_0 / k \sles\ 3 \times 10^3$~cm$^{-3}$~K
for $\Sigma({\rm H_2}) / \Sigma(\HI) \sles\ 0.5$ \citep{Yim2011}.
To strip the ISM from the satellite, the satellite must approach the
disk at a velocity 
\begin{eqnarray}
V_{sat} > 210 {\rm ~km~s}^{-1} 
\left( {P_{sat}} \over 3 \times 10^3 {{\rm ~cm}^{-3} {\rm ~K}}   \right)^{1/2} \\ \nonumber
\left( {4 \times 10^{-4} {\rm ~cm}^{-3}} \over {n_0} \right)^{1/2},
\end{eqnarray}
which is comparable to the circular velocity of \n891\ (Table
\ref{tab:ngc891}) 
and therefore a tenable explanation.

\begin{figure*}
\begin{center}
  \includegraphics[width=\linewidth]{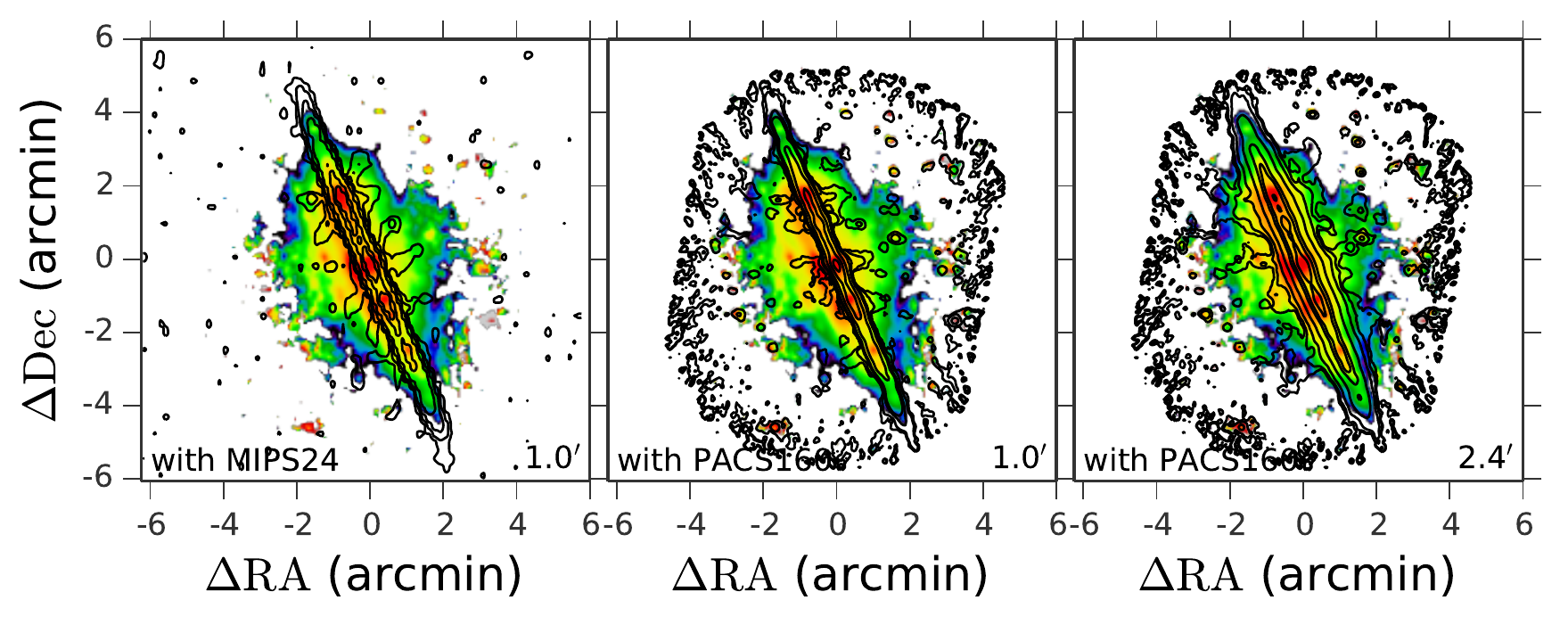}
\caption{The color map shown in Figure~\ref{fig:pacs_color} is shown with contours of 24\um\ (left) and  160\um\ (middle and right) after image sharpening (see text). }
\label{fig:sharp}
\end{center}
\end{figure*}

\subsubsection{Recent Accretion}

Our analysis suggests that recent accretion events produced both the dust spur and the \HI\ 
filament, but the nature of these events were different based on the distinct 
compositions of the two structures.
The interaction of an infalling gas-rich satellite or  gas stream with the CGM plausibly generated 
the \HI\ spur, but a collisionless component such as a dark matter minihalo was required to 
pull the dust spur out the back side of the disk.  These processes deliver gas to the
disk of NGC 891, thereby fueling further growth of the stellar disk.

Studies of the stellar halo of NGC 891 
show evidence that accretion of satellite galaxies has formed giant streams much like those mapped
 around the Milky Way \citep{Ibata2003,Yanny2003,Belokurov2007}.
\citet{Mouhcine2010} resolve structures that loop around the galaxy reaching heights of
$\approx 30$~kpc  much like the rosette-shaped pattern generated by a tidally disrupting dwarf galaxy.  
We note that the \HI\ spur follows
one of the stellar streams where it plunges toward the disk. This alignment is most easily seen in 
Figure~4 of \citet{Schulz2014} who identify seven faint satellites beyond this stellar stream. 
This satellite population provides more frequent interactions with the disk than the more massive
group member UGC~1807 (1:10 mass ratio) over $ 80$~kpc northwest of \n891. 
Interaction with a satellite may also explain the lopsidedness of the disk of 
\n891 \citep{Oosterloo2007a}.

\subsection{Origins of Halo Dust}  \label{sec:extrapolate_halo}

We have detected thermal emission from $1.5 \times 10^7$\msun\ of cool dust in the inner halo of \n891, i.e.,
the region within the ellipse drawn in Figure~\ref{fig:clean24}.  We estimate the total halo dust mass 
from an extrapolation of the halo surface brightness profile. 
We use the 24\um\ profile on the west side of the disk and
assume the halo extends vertically from a base of radius 20.9~kpc. We equate the central surface brightness of 
6.0 MJy~sr$^{-1}$ with a mass surface density of 0.20\msun\ pc$^{-2}$ based on Figure~\ref{fig:pix2pix},
arriving at  a total halo dust mass of $4 \times 10^7$\msun.

With this extrapolation, the halo dust mass increases from 15\% to roughly 43\% of the dust mass in the thick
disk.  The estimated dust mass in the halo therefore approaches the cosmic ratio of halo dust to disk dust.
\citet{Menard2012a} measured the reddening of background quasars caused by dust in \MgII~absorbers and
found that these absorbers account for most of the dust inside the virial radii of galaxies and that  
the galaxy halos contain at least 52\% as much dust as do galaxy disks. While we cannot rule out the 
possibility of substantially more halo dust in a cooler component at $T\sim7~\rm K$, the similarity of these 
numbers suggests that the {\it Herschel} imaging has detected the main component of halo dust in \n891.

We have identified two structures, a dusty superbubble and a dust spur, that may pollute the CGM with dust.
To add further insight into exactly how dust is transported into the CGM, 
we applied unsharp masking techniques to highlight the high spatial frequency structure
in the dust emission. Regardless of the exact procedure used, the results in Figure~\ref{fig:sharp} show 
dust filaments emanating from the hotspots associated with the molecular ring
as well as from the central hotspot. These filaments are detected over the full diameter of the molecular ring
and  extend roughly 4~kpc from the midplane. These dust filaments appear to be associated with the feedback
processes that are mixing dust into the halo; they are found over a larger region of the disk than 
the base of the superbubble. It should be possible (in future work) to compare the properties of such
filaments directly to wind simulations that take the interactions of the bubbles driven by multiple
star-forming regions into account \citep{Tanner2016}.

The halo dust clearly traces a fluid that is not pristine gas. While this material most likely originated in 
galaxies, the question remains whether it was primarily blown out of the disk by feedback processes or whether
tidal streams and ram pressure stripping of satellites have contributed a substantial fraction of the mass.
The newly discovered dust spur clearly shows that interactions with satellites do play some role.
The mass in the dust spur is about 25\% of that in the superbubble, however; and the filaments are also
lifting dust about 4 kpc off the disk midplane. Hence our results favor a dominant role for stellar feedback
in enriching the halo with dust.

To determine whether it is energetically plausible for feedback to lift most of the gas associated
with the halo dust out of the disk, we conservatively estimate the total mass of gas lifted into the halo
along with the dust. The dust-to-gas mass ratio of the galaxy as a whole is $M(dust)/M(gas) \approx 1 / 110$.
We estimate $M(gas) \approx 1.7 \times 10^9$\msun\ just above the disk and, very roughly for the entire halo, 
$M(gas) \approx 4 \times 10^9$\msun. A lower dust-to-gas ratio would yield more halo gas associated with the dust,
but the average galactic value is most appropriate for testing the hypothesis that the dust was lifted out of the
disk by feedback. We note that $4 \times 10^9$\msun\ is a small fraction of the total 
mass found in the cool CGM of galaxies \citep[$7-12\times10^{10}~\msun$]{Werk2014a}.

Following \citet{Howk1997a}, we assume an isothermal
sheet model for the vertical distribution of light and mass in \n891. Based on the similarity
of the rotation curves for \n891 and the Milky Way, we adopt the Milky Way midplane mass density $\rho_0$ 
\citep{Bahcall1984} as a proxy for \n891; the mass scale height $z_0$ is taken from infrared imaging
of \n891 \citep{Aoki1991}.
The potential energy, $W$, of a cloud  of mass $M_{\rm cl}$ a height $z$ above the midplane is then
\begin{eqnarray}
W  = 10^{52} {\rm~ergs} \left( {M_{\rm cl}}\over{10^5~\msun} \right) 
\left( {z_{\rm 0}}\over{700\rm~pc}\right)^2  \\ \nonumber
\left( {\rho_{\rm 0}}\over{0.185~\msun~\rm pc^{-3}} \right) 
{\rm ln} \left[ {\rm cosh} \left({z}\over{z_{\rm 0}} \right) \right].
\end{eqnarray}
The energy required to lift the $4 \times 10^9$\msun\ of \HI, $\rm H_2$, and dust in the superbubble
to a height of 4.3~kpc is roughly
\begin{eqnarray}
W \approx  1.2 \times 10^{58} {\rm ~ergs}
\left( {M_{\rm cl}}\over{4 \times 10^9~\msun} \right) 
\left( {z_{\rm 0}}\over{700\rm~pc}\right)^2  \\ \nonumber
\left( {\rho_{\rm 0}}\over{0.185~\msun~\rm pc^{-3}} \right). 
\end{eqnarray}
For a constant SFR of 4.8\msunyr, supernovae and stellar winds 
provide a mechanical power, $L_{\rm w} \approx 1.2 \times 10^{42}$~ergs~s$^{-1}$. 
Assigning an efficiency $\epsilon$ for transferring this energy to the cool gas and dust,
the time required to enrich the halo with dust is roughly 
\begin{eqnarray}
t \approx 3.1 \times 10^8 \epsilon^{-1} {\rm ~yr}. 
\end{eqnarray}
The current rate of star formation in the disk provides enough feedback energy to lift
the dust and associated gas into the halo in just a few rotational periods only if the 
feedback efficiency is very high. However, if we picture the CGM as continually recycling
disk material over a longer timescale, then feedback can easily account for most of halo dust in \n891.
For example, if the disk of \n891 was assembled at $z \sles\ 2$, and the SFR has steadily declined 
over the last 10.3~Gyr, then the dust can be lifted into the halo with inefficient feedback characterized by 
$\epsilon \approx\ 0.03$.






\subsection{Comparison to \n4631}


It is interesting to compare the feedback in \n891 to that in \n4631, another edge-on galaxy which we
recently studied with {\it Herschel} in Paper I. The galaxies \n4631 and \n891 have 
comparable \Ha, X-ray, and radio luminosities. Linearly polarized radio continuum emission has been 
detected in the halos of both galaxies and shows that magnetic field lines reach into the halos of both galaxies
\citep{Hummel1991}. The FIR to radio flux ratio for \n891 is $q_{60\mu m}=2.19$ in  Table~\ref{tab:fir_radio}. The FIR luminosity at 60\um\ is $9.64-9.70\lsun$ \citep{Sanders2003,Surace2004} and radio continuum luminosity is $21.85-21.76~\rm W~m^{-2}~Hz^{-1}$ \citep{Condon2002,Murphy2009}. Thus, we estimate $q_{60\mu m}=2.03-2.19$ for \n4631. This ratio is a sensitive probe of age 
during a starburst phase \citep{Bressan2002}, and  the comparable values for \n4631 and \n891 indicate
both galaxies are in a post-starburst phase.

In spite of these similarities, the halos of the two galaxies show some significant
differences. The density and scale height of the warm ionized gas are a bit larger
in \n4631, and the magnetic field extends further into the halo of \n4631 \citep{Hummel1991}.
The direction of the magnetic field lines is not well constrained for either galaxy due to
poorly constrained models for Faraday rotation, but after correction the foreground
Faraday rotation the field lines are quite different in the two galaxies.
The magnetic field lines point radially outward from the disk of \n4631 while 
no overall orientation is visible in \n891 where the field appears to be drawn 
into the halo by more local events (especially on the south side of the disk).\footnote{
        The radio continuum emission in Fig.~\ref{fig:pacs_color}
        also shows an extension in total intensity away from the disk plane to the southwest (opposite the dust spur). 
        The linear polarization of this southwest extension is larger than the polarization in any other region
        of \n891, and the polarized emission extends further into the halo here than it does above the superbubble
        \citep{Hummel1991}. 
        Bulk motion has convected the magnetic field outwards, and the adiabatic losses of the electrons
        may explain the changes in spectral index reported by \citet{Hummel1991}.  Whether there is a direct connection
        between this inferred velocity gradient and the dust spur remains unclear.}
Overall, these results suggest the galactic wind is stronger  in \n4631 than in \n891. 

Stronger feedback in \n4631 would be expected based on its higher SFR surface
density and gas fraction. 
{\jhy The SFRSD of \n4631 is about 5 times larger than that of \n891 as estimated in Section~\ref{superbubble_wind.sec}.}
It is also clear that \n4631 has a higher  gas fraction than \n891 based on its higher \HI\ mass, lower dynamical mass, and later spiral type 
\citep{rc3}.


It is therefore quite interesting to compare the dust properties of these two halos.
We find the {\jhy measured} dust mass of \n891 is 2.6 times higher than \n4631 in Paper I. 
We attribute this result to the larger size and mass of \n891. 
The superbubble in \n4631 is slightly smaller and contains a factor of two less dust mass
than the \n891 superbubble. 

The temperature of the dust, however, differs significantly between the two superbubbles.
The temperature of the dusty superbubble is significantly higher in \n4631, $T_{\rm dust} = 24.46 \pm 0.77$~K. 
When we include the 100\um\ data point from the HEDGES data, the superbubble temperature for \n891 
increases by $\sim 1$ K which is still 3~K below the  the temperature of the superbubble in \n4631.
When we take the 70\um\ data point as an upper limit, the temperature decreases slightly.
Therefore, the temperature difference of superbubble for \n4631 and \n891 cannot be alleviated by 
different SED fitting processes.\footnote{
      For the SED modeling, we used 5 bands, 70\um, 160\um, 250\um, 350\um,
      and 500\um\ while Paper I adopted 100\um\ in addition to
      the 5 bands.       Also, in Paper I,
      the 70\um\ data point was considered as an upper limit. 
      The higher temperature of the SED fit for \n4631 is driven
      by the data point at 100\um. 
      When we include the 100\um\ data point from
      the HEDGES data, the temperature of all components of \n891 increases by $\sim 1$~K.
}

Dust in galaxy halos can be heated by  the evolved stellar population 
or processes associated with the stellar feedback. Using B and V band photometry,
\citet{Ann2011} measured a scale height for the thick stellar disk of \n4631 of 
$1.33 \pm 0.25$~kpc and $1.25 \pm 0.28$~kpc, and \citet{Morrison1997} report an
R band scale height of $1.5-2.5$~kpc for \n891. It therefore seems unlikely 
that the stellar halo of \n891 is any less effective at heating dust than 
that in \n4631.
The SFR surface density of \n4631 is only slightly higher than that of \n891,
so an attempt to explain the difference in superbubble temperatures would need to appeal to
the lower dust mass and smaller size of the superbubble in \n4631. It would be interesting
to explore whether these properties of the \n4631 superbubble reflect a younger age or 
the more ordered magnetic field of the halo. The increased temperature of the superbubble might
reflect the larger pressure required to push these field lines out of the superbubble's path.



\section{Conclusions} \label{sec:conclusions}




We present the results of our {\it Spitzer}/MIPS and {\it Herschel}/PACS
GO observations of \n891. The  angular resolution and
sensitivity allow us to resolve the distribution of dust in the disk and halo. 
We place these components in the context of the stellar component and multi-phase ISM 
using supplementary observations from the radio to the X-ray. We obtain the
following insight into the disk - halo interaction.

\begin{itemize}

\item The MIPS image at 24\um\ shows a dust spur extending 20 kpc southeast of the galaxy.
The PACS and SPIRE images confirm the presence of this feature. Fitting a MBB model to the
SED indicates a dust mass of $7.6 \times 10^5$\msun, which is surprisingly large compared
to the upper limits on the \HI\ gas mass. The high dust-to-gas ratio is inconsistent with a
primordial origin (e.g., a cold stream of infalling material) and indicates the dust was
likely pulled out of \n891 by a satellite interaction. Its composition differs distinctly
from the \HI\ spur to the northwest where we do not detect thermal dust emission. 
The local maxima in the far-infrared surface brightness coincide with knots of radio
continuum emission, perhaps a sign that they are young star-forming regions. The absence
of optical or near-infrared emission from the dust spur suggests the colliding satellite
had little stellar mass, suggesting that the dust spur may be the remnant of a minihalo
collision with the disk.


\item 
The PACS 70/160\um\ ratio map draws attention to a superbubble extending over 7 kpc above
the midplane on its western side. We find a close correspondence between the superbubble and
the thermal X-ray emission suggesting that the thermal pressure of hot gas drives the expansion. 
We argue that the superbubble is likely accelerating as it breaks
through the disk because its height exceeds the pressure scale height of not only the thin
disk but also the gaseous halo. The feedback associated with the star formation activity 
within the inner disk appears to be sufficient to launch this galactic wind. The SFR surface
density of 0.03\msunyr\ kpc$^{-2}$ and gas fraction ($\sles\ 10\% $) suggest somewhat lower
threshold values for driving winds than do the values measured previously in starburst galaxies.

\item
This disk contains regions of hot dust associated with the galactic center and the molecular ring.
Image sharpening techniques reveal dust filaments rising roughly 4 kpc above these regions. The
dust temperature is higher in the inner disk than the halo, and the mass surface density of the
dust is centrally concentrated. 

\item
{\jhy The halo dust component is detected 10~kpc or more above the entire stellar disk with the measured halo dust mass of $M_{\rm dust,halo} = 1.5\times10^{7}$\msun.} A rough extrapolation of the dust mass surface
density suggests $\approx 4 \times 10^7$\msun\ of dust has been lifted into the halo carrying
with it a gas mass roughly 100 times larger. This mass flux accounts for roughly 10\% 
of the gas mass estimated to reside in the CGM of similar galaxies.

\end{itemize}

Because \n891 is a very close analog of the Milky Way in respect to many of its properties, 
we believe these results provide a viable picture of how dust is deposited in the halos of
typical galaxies. The elevated SFR and molecular gas fraction with respect to the Milky
Way would perhaps most closely describe the Milky Way in its recent past when its SFR
was 
higher. In this sense, \n891 provides a direct image of the
emission from dust and multi-phase gas in the inner halo of a galaxy where the baryon
budget of the CGM is well constrained observationally \citep{Hodges-Kluck2018,Das2020}.

\section*{Acknowledgements}

We thank Cassi Lochhaas for discussions about the circumgalactic bubble which
clarified how its structure depends on SFR surface density,
and we thank an anonymous referee 
for suggesting we apply the Dahari parameter.
This work was supported in part by the National Science Foundation (NSF) under
AST-1817125 (CLM) and ASTR1009583 (SV), JPL Awards 1276783 and
1434779, and NASA grants NHSC/JPL RSA 1427277, 1454738
(SV and MM), and  ADAP NNX16AF24G (SV).  Some of the work was completed
at the Aspen Center for Physics,  and accompanying support from the NSF through
PHY-1066293 is gratefully acknowledged.  
We deeply regret that the co-author (CE) who provided the
\Ha\ image and made a significant contribution to the analysis of the
{\it Spitzer} data did not live to see the article published.
{\it Herschel} is an ESA
    space observatory with science instruments provided by
    European-led Principal Investigator consortia and with important
    participation from NASA.


\section*{Data Availability}

The data underlying this article will be shared on reasonable request
to Dr. Joo Heon Yoon.




\bibliographystyle{mnras}
\bibliography{references_new} 

\bsp	
\label{lastpage}
\end{document}